\UseRawInputEncoding
\documentclass[twocolumn,secnumarabic,amssymb,nobibnotes,aps,prl]{revtex4-1}

\usepackage{graphicx}% Include figure files
\usepackage{dcolumn}% Align table columns on decimal point
\usepackage{bm}% bold math
\usepackage{multirow}
\usepackage{mathptmx}%英文字（包括数学公式中的英文）字体为New Times Roman
\usepackage{amsmath}%高级数学模式宏包
\usepackage{amssymb}%高级数学符号宏包
\usepackage{bibentry}
\usepackage{setspace}
\usepackage{mathtools}
\usepackage{color}
\usepackage{CJK}
\usepackage{float}
\usepackage[CJKbookmarks,colorlinks,linkcolor=blue]{hyperref}
\usepackage{url}

\usepackage{graphicx}% Include figure files
\usepackage{dcolumn}% Align table columns on decimal point
\usepackage{bm}% bold math
\usepackage{multirow}
\usepackage{mathptmx}%英文字（包括数学公式中的英文）字体为New Times Roman
\usepackage{amsmath}%高级数学模式宏包
\usepackage{amssymb}%高级数学符号宏包
\usepackage{bibentry,natbib}
\usepackage{setspace}

\begin{document}
\title{Classical, semi-classical and quantum optical models for x-ray planar cavity with electronic resonance}%

\author{Xin-Chao Huang$^{1\dag}$\footnote{xinchao.huang@xfel.eu}, Tian-Jun Li$^{2}$\footnote{Authors contribute equally}, Frederico Alves Lima$^1$ and Lin-Fan Zhu$^2$\footnote{lfzhu@ustc.edu.cn}}

\affiliation{$^1$European XFEL, Holzkoppel 4, 22869 Schenefeld, Germany\\
$^2$Department of Modern Physics, University of Science and Technology of China, Hefei, Anhui 230026, People's Republic of China}

\begin{abstract}
Here two theoretical models of semi-classical matrix and quantum Green's function are developed for the system of x-ray thin-film planar cavity with inner-shell electronic resonances. The semi-classical model is based on the matrix formalism to treat each layer as the propagating matrix. The crucial idea is to expand the propagating matrix of the resonant atomic layer under ultrathin-film approximation, then derive the analytical expression of the spectral observation, i.e, the cavity reflectance. The typical cavity effects of cavity enhanced decay rate, cavity induced energy shift and the Fano interference which were observed in the recent experiments could be phenomenologically interpreted. The second quantum model employs the analytical Green's function to solve the cavity system. The system Hamiltonian and the effective energy-level are derived. The effective energy-level scheme indicates that the cavity effect acts on the regulation of the intermediated core-hole state. To test the validity of the semi-classical matrix and quantum Green's function models, the classical Parratt's formalism and the dispersion correction of the atomic refractive index are also recalled. Very good agreements in reflectivity spectra between semi-classical and quantum models with the Parratt's results are observed. The present semi-classical matrix and quantum Green's function models will be useful to predict the new phenomena and optimize the cavity structures for future experiments and promote the emerging of quantum optical effects with modern x-ray spectroscopy techniques.

\end{abstract}

\date{\today}%

\maketitle

%\tableofcontents
%\begin{spacing}{2.0}

\section{I. Introduction}
X-ray quantum optics has become a flourishing field of research in last decade~\cite{adams2013x,rohlsberger2021quantum,kuznetsova2017quantum,wong2021prospects,ueda2019roadmap,rohringer2019x} with the developments of the x-ray sources such as high-brilliant synchrotron radiation (SR)~\cite{PhysRevAccelBeams.24.110701} and x-ray free electron laser (XFEL)~\cite{RevModPhys.88.015007}, x-ray detection and sample fabrication. Depending on the x-ray intensity, non-linear and linear quantum phenomena have been extensively studied by XFEL~\cite{young2010femtosecond,chumakov2018superradiance,Fukuzawa2013,PhysRevLett.127.213202,Tamasaku2018,yoneda2015atomic,PhysRevLett.123.023201,PhysRevLett.117.027401,PhysRevLett.121.137403,PhysRevA.104.L031101} and SR~\cite{PhysRevLett.95.097601,Rohlsberger1248,rohlsberger2012electromagnetically,PhysRevLett.111.073601,PhysRevLett.114.207401,PhysRevLett.114.203601,haber2017rabi} respectively. Hard x-ray XFELs usually provide individual pulses with $\sim10^9$ photons/eV with duration on the order of a few to tens of femtoseconds~\cite{young2010femtosecond}, thus being ideal for studies of non-linear phenomena such as multi-photon excitation of Dicke superradiance~\cite{chumakov2018superradiance} for nuclei isotope, and multiphoton ionization~\cite{Fukuzawa2013} and resonant excitation~\cite{PhysRevLett.127.213202}, two-photon absorption~\cite{Tamasaku2018}, population inversion~\cite{yoneda2015atomic}, and superfluorenscence\cite{PhysRevLett.123.023201}, stimulated emission~\cite{PhysRevLett.117.027401,PhysRevLett.121.137403,PhysRevA.104.L031101} and etc for inner-shell electronic resonances.
The peak intensity of SR pulses is 5-6 orders of magnitude lower than those of XFELs, and their pulse duration is $\sim$3 orders of magnitude longer, therefore up to now almost all experiments done in SR were restricted to the linear or so called weak excitation regime. Nevertheless, a vast number of fundamental studies about collective, interference and cavity effects using SR have been reported~\cite{PhysRevLett.95.097601,Rohlsberger1248,rohlsberger2012electromagnetically,PhysRevLett.111.073601,PhysRevLett.114.207401,PhysRevLett.114.203601,haber2017rabi}. Wherein, a very successful physical platform is the x-ray thin-film cavity or called x-ray cavity QED setup with a usual thickness of nanometers, in which a rich variety of quantum optical phenomena have been realized via it, such like lifetime shortening~\cite{PhysRevLett.95.097601}, the collective Lamb shift~\cite{Rohlsberger1248}, electromagnetically induced transparency~\cite{rohlsberger2012electromagnetically}, spontaneously generated coherences~\cite{PhysRevLett.111.073601}, Fano interference~\cite{PhysRevLett.114.207401}, group velocity control for x-ray photons~\cite{PhysRevLett.114.203601}, the collective strong coupling of x-rays and nuclei~\cite{haber2017rabi}. The observation of collective Lamb shift is a particular milestone in the field of x-ray cavity-QED science, which inspires the similar researches in optical regime using atomic ensembles~\cite{PhysRevLett.108.173601,PhysRevLett.113.193002,PhysRevLett.117.073003}, demonstrating the coherent control ability of the x-ray cavity and breading the new field of x-ray quantum optics.

So far most of cavity QED studies above were based on the M\"{o}ssbauer nuclear resonances. The cavity can be regarded as a structure with boundary limitation for electromagnetic field~\cite{ivchenko2013resonant}, and in general, the role of cavity is to modify the electromagnetic environment and the mode density. Therefore the coupling between the atoms/nulei and the photons will be remarkably strengthened, resulting in the typically observed phenomena of line broadening and energy shift~\cite{RevModPhys.73.565}. In hard x-ray regime, another similar system of photonic crystal provides possible evidence to the reasons of the cavity effect with electronic resonance being so rarely noticed in previous studies. The photonic crystal is a periodic quantum-well multilayer structure that can also modify the mode density around the Bragg peak, but it only results in the angular modulations of x-ray spontaneous fluorescence with electronic resonators which is known as Kossel effect~\cite{kossel1935richtungsverteilung} without obvious linewidth broadening and energy shift~\cite{jonnard2002modulation,andre2015kossel,andre2010x}. However, a more anomalous behaviours of collective strong coupling and anti-crossing could be easily realized with the resonator of nuclei~\cite{haber2016collective,andreeva2018nuclear}. The coupling strength between the atom and cavity mode is related to the atomic dipole moment~\cite{RevModPhys.73.565}, and following the x-ray scattering framework we know that the resonant scattering length which depends on the dipole moment of the nuclear transition is 1-2 orders of magnitude larger than the one of inner shell transition~\cite{rohlsberger2004nuclear}. Therefore coupling strength will be stronger and the cavity effect could be more obvious in nuclear resonance. Nevertheless, a much stronger modification of the mode density can be provided by the x-ray thin-film cavity compared with the periodic multilayer structure, as demonstrated in recent studies exploiting the viability of inner shell electronic resonances in x-ray thin-film cavities~\cite{Haber2019,PhysRevResearch.3.033063,vassholz2021observation}. The lifetime and energy of the core-hole state are simultaneously modified in the thin-film cavity~\cite{PhysRevResearch.3.033063}, which are manifestations of the typical Purcell and collective effects leadings to controllable spcetral shape in materials~\cite{Haber2019}. Vassholz and Salditt reported the observation of directional spontaneous emission of K$\alpha$ emission lines of cobalt, copper and iron~\cite{vassholz2021observation}. It is important to note that the directional spontaneous emission is also a well-known cavity effect that has been extensively studied at optical regime~\cite{PhysRevA.41.2668,ujihara1993decay}. Moreover, Gu \emph{et al} subsequently proposed broad prospects for molecular core-hole state manipulation with the x-ray cavities, such like creating core polaritons and delocalizing the core-hole states~\cite{PhysRevLett.126.053201,gu2021manipulating}. These pioneer works indicate that the x-ray cavity with electronic resonances will become a new promising platform for x-ray quantum optics, and promote the merging between x-ray technology progress and quantum optics effects.

\begin{figure}[htbp]
\centering
\includegraphics[width=0.45\textwidth]{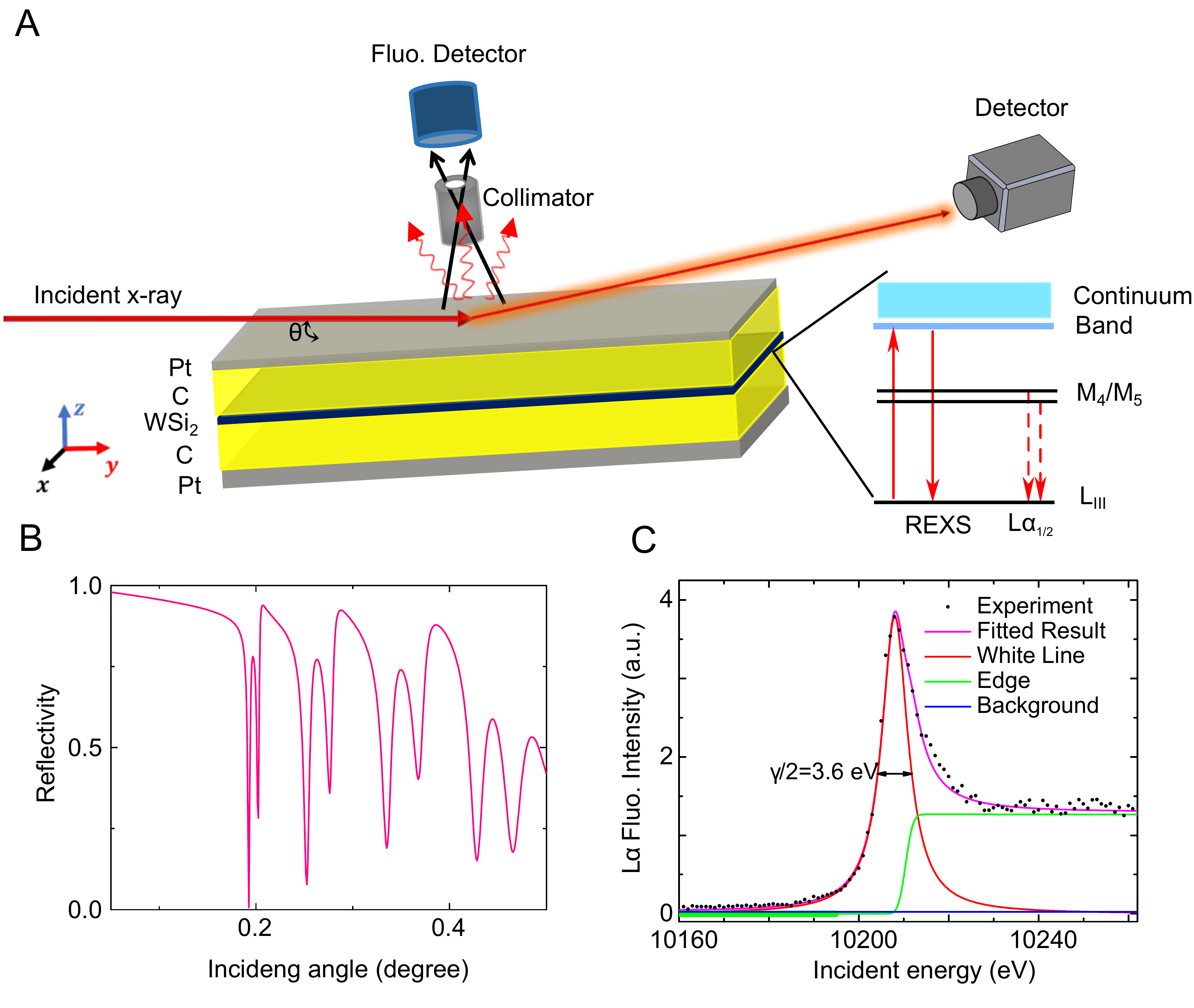}
\renewcommand{\figurename}{Fig.}
\caption{\label{Fig.1}The schematics for x-ray thin-film planar cavity and the measurement setup usually used in experiment. (A) Cavity structure. The cavity has a sandwich structure of cladding layers, guiding layers and atomic layer. In this present work, a geometry of Pt (2.1 nm)/C (28.2 nm)/WSi$_2$ (2.0 nm)/C (28.2 nm)/Pt (16.0 nm)/Si$_{100}$ is chosen. The middle inset shows the energy-level of L$_{\textrm{III}}$ edge of atom W, the transition energy of the white line is 10208 eV. After the excitation from 2$p_{3/2}$ to 5$d$ band via x-ray photon, there are competitive decay channels that is the particular character of the inner shell excitation. The solid arrow represents the resonant elastic scattering process which is well-known as resonant elastic x-ray scattering (REXS), and the dash arrows represent the inelastic scattering, which could be the fluorescence emission or resonant inelastic x-ray scattering (RIXS). The Auger processes are not depicted here. (B) The calculated $\theta-2\theta$ rocking curve for the bare cavity without considering the atomic resonance. The reflection dips of the bare cavity correspond to the specific cavity modes, in present work the incident angles are chosen around the first dip, i.e., the first cavity mode for showing and comparing the results. (C) An example of the experimental and fitted fluorescence spectra~\cite{PhysRevResearch.3.033063} as a function of incident photon energy at an incident angle of 3$^\circ$. The solid dots are experimental results, and the solid lines in pink, red, green and blue are the fitted result, Lorentzian resonance line, ionization continuum line and the flat background respectively. The fluorescence spectrum will be used to derive the refractive indexes of the WSi$_2$ material with resonance contribution.}
\end{figure}

Fig. \ref{Fig.1}(a) depicts a typical example of the cavity structure, which is made of a multilayer of cladding and guiding layers. The cladding layers act as mirrors with high electron density materials such as Ta, Pd and Pt, while the guiding layers stack the cavity space with low electron density materials such as $\rm{B_4}$C, B and C. To guarantee the cladding layer has a good reflection ability with avoiding strong absorption loss, the thin-film are used at grazing incident geometry in general (normally the incident angle is below the critical angle of the cladding layer) because the refractive index of all materials at x-ray energies is always smaller than unity. The top mirror layer is relatively thin so that the x-ray can be evanescently coupled into the cavity. This kind of cavity structure is also called as x-ray 1D waveguide which is usually used to focus the x-ray beam to nanometers scale and enhance the fluorescence yield~\cite{PhysRevLett.109.233907,PhysRevLett.100.184801,PhysRevB.64.233403}. In this design, at certain incident angles $\theta_{\textrm{th}}$, x-ray can resonantly excite specific cavity guided modes where the reflection dips in the rocking curve appear as shown in Fig. \ref{Fig.1}(b). Then the coupling between the cavity and atom is built by embedding a thin atomic layer inside the cavity.

From the classical point of view, the reflectivity and the scattering field of the multilayer structure can be well solved by the scattering theories, such as the kinematical approximation~\cite{Kinematicalapproximation}, coupled-wave theory and Parratt's formalism~\cite{PhysRev.95.359,windt1998imd}. To solve the scattering processes, the central key is the accurate refractive indexes of all multilayer materials. For the cladding and guiding layers, the incident energy is usually far away from their absorption edges, so the refraction indexes are flat and energy-independent. For the atomic layer with resonances, the refraction index is energy-dependent which is commonly called anomalous resonant scattering~\cite{fink2013resonant,RevModPhys.93.035001,abbamonte2002structural}. It should be clear that after correcting the dispersion by the resonant scattering length under weak excitation regime, there is nothing really anomalous~\cite{als2011elements}. By accurately translating the atomic or nuclear resonances into the energy-dependent refractive indexes, the Parratt's formalism is still adequate to calculate the spectral behaviors of the cavity structure. In both of electronic~\cite{PhysRevB.44.498,PhysRevB.98.235146,yang2013making,tiwari2010applications} and nuclear~\cite{rohlsberger2004nuclear,rohlsberger1999theory,sturhahn2000conuss,andreeva2008nuclear,huang2017field} resonances systems, the Parratt's formalism is a benchmark for developing novel semi-classical and quantum optical models in nuclear system because it can provide accurate spectrum for the cavity reflectivity which is in agreement with the experimental results. The Parratt's formalism combined with the reciprocity theorem provides a method to calculate the x-ray fluorescence spectrum for electronic resonances in cavity structures~\cite{andre2015kossel,li2014correction}. However, using solely the classical model is not possible to provide an interpretation of the quantum optical effects. The experimental spectra cannot be correlated with the retrieved physical parameters too.

The Parratt's recursion can be actually remoulded in the form of transfer matrix, which has been successfully implemented in calculation of nuclear resonance~\cite{rohlsberger2004nuclear,PhysRevB.69.235412}. When the thickness of the nuclear layer is as thin as to nanometer scale, the matrix of the resonant part could be treated by Taylor expansion and the high-order contributions be omitted. In this case, the scattering field amplitude from the bare cavity, i.e., without the nuclear resonance, will be separated from the one of resonant part, then an analytical reflection coefficient with line broadening and energy shift effects could be derived. Using such semi-classical model, the line broadening could be phenomenologically interpreted as the decay rate speedup~\cite{PhysRevLett.95.097601,andreeva2008nuclearNIMB}, and the energy shift is associated with the collective Lamb shift~\cite{Rohlsberger1248}. In the recent work of controlling core hole lifetime of the inner shell electronic resonance, the semi-classical model was also used to explain the linear relation between the enhanced decay rate and the field amplitude inside the cavity~\cite{PhysRevResearch.3.033063}. However, there are some limitations of the semiclassical matrix method, for example the formalism will be very complex for the cavity with embedding multiple atomic or nuclear layers~\cite{rohlsberger2012electromagnetically,Huang2020} or double-cavity structure~\cite{haber2017rabi}, and it is also hard to derive the effective energy-levels which is crucial for correlating the experimental spectral observations with physical parameters. Therefore, with the development of the x-ray cavity-QED with the nuclear resonance in last decade, several quantum optical models have been well developed for interpreting and predicting the experimental phenomena, including the phenomenological quantum optical model~\cite{PhysRevA.88.043828,PhysRevA.91.063803} based on the Jaynes-Cummings approach~\cite{jaynes1963comparison}, the \emph{ab initio} few-mode quantum model~\cite{PhysRevResearch.2.023396,PhysRevX.10.011008,lentrodt2021classifying,diekmann2021inverse} and the \emph{ab initio} Green's functions model~\cite{PhysRevResearch.2.023396,PhysRevA.102.033710,PhysRevA.104.033702}.

Studies of x-ray cavity with electronic resonances are more recent than those reporting on nuclear ones, thus there is currently a gap in effective theoretical tools dealing with quantum optical framework. The very recent experimental works~\cite{Haber2019,PhysRevResearch.3.033063,vassholz2021observation,PhysRevLett.126.053201,gu2021manipulating} also encourage future studies with more complex cavity structures and modern x-ray spectroscopy techniques, which need a versatile theoretical quantum model. Inspired by the well-established theoretical models in the nuclear resonance system, here we extend the semi-classical matrix and the quantum Green's function models of the x-ray thin-film planar cavity into the electronic resonance system. We benchmark these two models by using the Parratt's calculation, the comparisons of the reflectivity spectra between them show very good agreement. Both two models can be successfully implemented to describe the recent experimental observations~\cite{Haber2019,PhysRevResearch.3.033063}, including the cavity enhanced decay rate (CER), the cavity induced energy shift (CIS) and the Fano interference. Especially, the Green's functions could be used to fit the dipole moment of the electronic resonance and derive the effective energy-levels following the language of quantum optics, and it can be conveniently extended to the periodic multilayer photonic crystal structures~\cite{PhysRevA.102.033710}, which will be useful to open new avenue of x-ray cavity studies in tender and soft x-ray regimes~\cite{PhysRevLett.126.053201,gu2021manipulating}, wherein the dipole moments are usually stronger than the ones in hard x-ray regime. These two models will be helpful to predict the new phenomena and design the cavity structure for future experiments.

The manuscript is organized as follows. Section I is the introduction. In section II, we recall the classical model and introduce the Parratt's formalism. The resonant elastic scattering amplitude corrections $\Delta f'(\omega)$ and $\Delta f''(\omega)$ are briefly discussed which are the crucial physical quantities for dispersion correction of the atomic refractive index. Reciprocity theorem is also applied to calculate the fluorescence spectra as functions of incident energy and angle, and the behaviors of line profile broadening and shift are observed. In section III, we introduce the semi-classical matrix method, utilizing the approach of atomic layer matrix expansion under thin film approximation, and the analytical formula of the reflection coefficient is obtained. The CER and CIS are connected with the real and imaginary parts of the field amplitude $\eta$ inside the cavity. The numerical reflectivity spectra agree well with the ones of Parratt's calculation, and can give the interpretations of two-pathway Fano interference and anti-crossing phenomenon. In section IV, we introduce the quantum Green's function model then derive the effective Hamiltonian and obtain the two key physical parameters of decay rate and spin-exchange. The Green's function in the structured media geometry was analytically derived by Toma\^{s}~\cite{PhysRevA.51.2545}, and we follow the same procedure in reference~\cite{PhysRevA.102.033710} to treat the thin-film cavity as 1D structured media. The input-output relation is employed to give the observed reflectivity spectra to compare with the Parratt's results, and very good agreements are shown. The value of the dipole moment is obtained via fitting process, which is compared with the one of nuclear resonance. Section V gives the summaries of the developed theoretical models. Finally we predict some possible applications based on the semi-classical and Green's functions models.

\section{II. Classical Parratt's Formalism}

\subsection{A. Basic Parratt's formalism}

L. Parratt extended the exact calculation of single layer to the multilayer and developed the Parratt's formalism~\cite{PhysRev.95.359}. It supposes that the multilayer is sitting on an infinitely sbustrate~\cite{als2011elements}, which holds the truth that the substrate in millimeter scale is much thicker than the mulitilayer structure in nanometers. Parratt's formalism has been proved to be the standard method to calculate the reflectivity curve and the field distribution inside the multilayer~\cite{huang2017field}. The cavity is indeed a particular sandwich structure of multilayer, therefore a numerical implementation of Parratt's formalism is still suitable. Firstly we give the basic expressions of the Parratt's formalism.

\begin{figure}[htbp]
\centering
\includegraphics[width=0.45\textwidth]{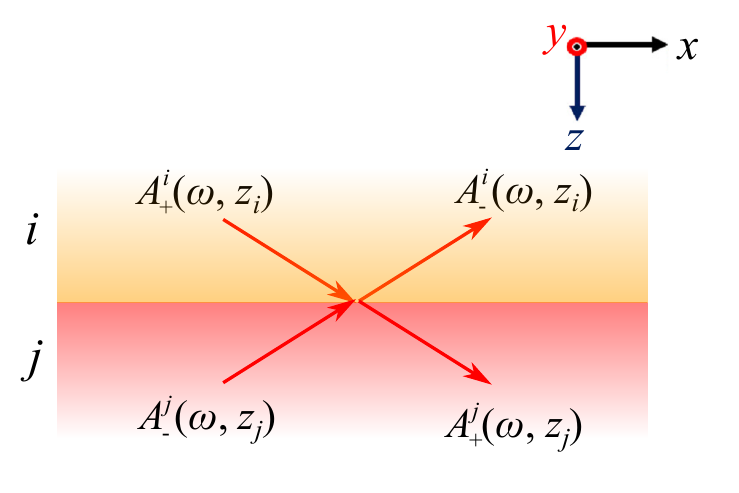}
\renewcommand{\figurename}{Fig.}
\caption{\label{Fig.2}The schematic of the Parratt's recursion. At the interface between the layers $i$ and $j$, the transmission and the reflection parts of the field are depicted as $+$ and $-$ directions.}
\end{figure}

The basic geometry of two layers with different thicknesses and materials is shown in Fig. \ref{Fig.2}, for the 1D structured media all of the layers and the interfaces between them are parallel. Inside a layer (e.g., located in the layer $i$), the field amplitude in layer $i$ at a depth $z_i$ is given by considering propagating of a plan x-ray wave with frequency $\omega$
\begin{equation}
A^i(\omega, z_i)=A^i(\omega) \times \textrm{exp}(ik_{i}\cdot z_i) \textrm{ ,}
\end{equation}Inside the medium $i$, the component of the wave vector $k_i$ in $z$ direction is written as
\begin{equation}
k_{i}=k \times \sqrt{{n_i}^2- {\cos}^2(\theta)}\textrm{ ,}
\end{equation}where $k$ is the wave vector of the incident x-ray in vacuum, $n_i$ is the refractive index of layer $i$, $\theta$ is the incident grazing angle. At the boundary between the medium $i$ and $j$, the complex refection efficient $r_{ij}$ and transmission coefficient $t_{ij}$ are given by the Fresnel equations
\begin{equation}
\begin{array}{lll}
r_{ij} = \frac{k_i-k_j}{k_i+k_j} \\
t_{ij} = \frac{2k_i}{k_i+k_j} \textrm{ ,}
\end{array}
\end{equation}note here that the polarization effect is left out. Then the reflection and the transmission field amplitudes could be derived by the Parratt's recursion
\begin{equation}
\begin{array}{lll}
A_ - ^i\left( {\omega ,{z_i}} \right) = \beta _i^2{\alpha_i}A_ + ^i\left( {\omega ,{z_i}} \right) \\
A_ + ^j\left( {\omega ,{z_j}} \right) = \frac{{{\beta _i}{t_{ij}}A_ + ^i\left( {\omega ,{z_i}} \right)}}{{2\left( {1 + \beta _j^2{\alpha _j}{r_{ij}}} \right)}}  \textrm{ ,}\\
\end{array}
\end{equation}with
\begin{equation}
\begin{array}{lll}
{\alpha _i} = \frac{{{r_{ij}} + \beta _j^2{\alpha _j}}}{{1 + \beta _j^2{\alpha _j}{r_{ij}}}} \\
{\beta _i} = \exp \left( { - i{k_i}{z_i}} \right) \textrm{ .}
\end{array}
\end{equation}The interface roughness is omitted here.

For a thin-film multilayer, there are two boundary conditions can be used to start the recursion. Firstly, the topmost layer $i=0$ is the air or vacuum media, in this case $A_ + ^0(\omega)=A^0(\omega)$ in which $A^0(\omega)$ is the incident x-ray field with frequency $\omega$. Secondly, the substrate layer $i=s$ is infinitely thick, so there is no field enters the structure from bottom, i.e., $A_ - ^s(\omega)=0$. Applying Eq. (4), the reflection and transmission field amplitudes at any depth $z$ could be obtained, which could be used to obtain the normalized field amplitude

\begin{equation}
a(\omega, z)=\frac{A_+(\omega, z)+A_-(\omega, z)}{A^0(\omega)} \textrm{ ,}
\end{equation}and the normalized field intensity

\begin{equation}
I(\omega, z)=|a(\omega, z)|^2 \textrm{ .}
\end{equation}The reflectivity of the overall multilayer is given by the ratio of the reflected field amplitude and the incident x-ray strength at the surface layer

\begin{equation}
r(\omega, \theta) = \frac{A_- ^0(\omega, \theta)}{A^0(\omega, \theta)} \textrm{ .}
\end{equation}Then the reflectance $|r|^2$ could be used to compare with the experimental measurement.

It can be seen that the Parratt's formalism can obtain the reflectance at any incident energies and angles for the multilayer, and the refractive indexes are the crucial parameters in the procedure of calculation. Firstly we consider the bare cavity structure without the atomic resonance. The working energy around 10 keV~\cite{Haber2019,PhysRevResearch.3.033063} is far away from the ionization thresholds of the cladding and guiding mediums, and their elastic scattering process with the x-rays is mainly from the contribution of Thomson scattering, therefore the refractive indexes are flat and energy-insensitive within a range of hundred electron volts. Subsequently to include the atomic resonance, the resonant elastic scattering of the atomic layer must be considered to correct the refractive dispersion.

\subsection{B. Resonant X-ray scattering}
The refractive index describes a fact that the transmission or reflection of x-ray is from a beam scattering when it passes through a slab of the material which consists a set of 'scatterers', and the total field is a superposition of the incident and scattering waves. When we consider a homogenous medium with identical atom scatterers, the refractive index is connected with the atomic scattering length or so called oscillator strength $f$~\cite{attwood2000soft,fink2013resonant,xu2015realization}

\begin{equation}
n=1-\frac{2\pi\rho_ar_0}{k^2}\times f \textrm{ ,}
\end{equation}wherein $r_0$ is the classical radius of the electron. $\rho_a$ is the number density of the material which can be calculated by the lattice parameters, e.g., for tungsten disilicide~\cite{lukovic2017tungsten} in the tetragonal space group gives $\rho_a=12.42$ nm$^{-3}$. When we talk about the scattering, it refers to the elastic processes. In general, there are two elastic scattering channels of the Thomson scattering and resonant elastic scattering. The second item describes the picture that the inner-shell electron firstly is excited to a higher intermediate state ${\left| e \right\rangle}$, i.e, the core hole state, then decays back to the initial ground state ${\left| g \right\rangle}$ through emitting a photon whose energy is same as the incident one. The overall differential cross-section of the elastic scattering is given by the famous Kramers-Heisenberg formula~\cite{RevModPhys.83.705,gel1999resonant,kramers1925streuung}

\begin{align}
\frac{{d\sigma }}{{d\Omega }} &= \nonumber \\
&r_0^2{\left| {\left\langle g \right|\rho \left( Q \right)\left| g \right\rangle  - m/\hbar \sum\limits_e {\frac{{\left\langle g \right|\hat J\left( {k'} \right)\left| e \right\rangle \left\langle e \right|\hat J\left( { - k} \right)\left| g \right\rangle }}{{{\omega _i} - {\omega _g} - \omega  - i\Gamma /2}}} } \right|^2} \textrm{ .}
\end{align}The polarization effect is omitted. Here the first item describes the Thomson scattering where $\rho(Q)$ is the charge density of atom in momentum space. The second item is a second-order matrix element which describes the resonant elastic scattering, and it relates to the virtual process that the electron is excited from ${\left| g \right\rangle}$ to ${\left| e \right\rangle}$ then decays back to ${\left| g \right\rangle}$. We need to note here that the second-order scattering can also happen via the inelastic fluorescence, inelastic scattering and Auger processes~\cite{Auger1925,Krause1979}, in these cases the final states ${\left| f \right\rangle}$ will be different from the ground state ${\left| g \right\rangle}$. These channels are competitive, for example the resonant elastic scattering and the L$\alpha$ processes are depicted in the inset of Fig. \ref{Fig.1}(a). ${\hat J}$ is the current operator, $k(k')$ is the incident (scattering) wave vector and $Q=k'-k$ is the momentum transfer. $\Gamma$ is the lifetime of ${\left| e \right\rangle}$ which could be accelerated by the cavity effect~\cite{Haber2019,PhysRevResearch.3.033063}.

Note that $\frac{{d\sigma }}{{d\Omega }} = {\left| f \right|^2}$, therefore the scattering length $f$ could be written as a summarization of two parts

\begin{equation}
f=f_T+\Delta f \textrm{ ,}
\end{equation}where $f_T$ is the Thomson scattering length, and $\Delta f$ is the dispersion correction from the resonant scattering. We emphasize that Eq. (11) is also suitable for the nuclear resonance, e.g., the transition energy of $^{57}$Fe is 14.4 keV which is far away from the K edge of Fe at 7.1 keV, therefore $\Delta f_N$ is only from the nuclear resonance and $f_T ^N$ is mainly from the Thomson scattering of iron atom~\cite{rohlsberger2004nuclear}. From this viewpoint of scattering theory, there is no difference to treat the nuclear or electronic resonant scattering in frequency space, hinting that the x-ray cavity effect should be equally applied to the electronic system. $\Delta f$ could be calculated by summarizing all intermediate states~\cite{PhysRevB.58.3741,PhysRevLett.77.1508,RevModPhys.83.705}, or by deriving from the experimental spectrum. Actually, the imaginary part of $\Delta f$ represents the absorption strength~\cite{fink2013resonant,sakurai2006advanced}

\begin{equation}
\frac{d\sigma _{abs}}{d\Omega}=\frac{4\pi}{k}\textrm{Im}[\Delta f] \textrm{ ,}
\end{equation}wherein $\frac{d\sigma _{abs}}{d\Omega}$ could be gotten from the experimental x-ray absorption spectrum (XAS), e.g., the fluorescence yield XAS of WSi$_2$ around L$_\textrm{III}$ is shown in Fig. \ref{Fig.1}(c). Then the real part of the complex $\Delta f$ could be derived from the imaginary part via the well-known Kramers-Kronig relations~\cite{kronig1926theory,kramers1927diffusion}

\begin{align}
{\mathop{\rm Re}\nolimits} \left[ {\Delta f (\omega)} \right] &= \frac{2}{\pi}{\mathcal{P}}\int_0^\infty  {\frac{{\omega '{\mathop{\rm Im}\nolimits} \left[ {\Delta f\left( {\omega '} \right)} \right]}}{{{{\omega '}^2} - {\omega ^2}}}d} \omega ' \nonumber \\
{\mathop{\rm Im}\nolimits} \left[ {\Delta f (\omega)} \right] &=  - \frac{{2\omega }}{\pi }{\mathcal{P}}\int_0^\infty  {\frac{{{\mathop{\rm Re}\nolimits} \left[ {\Delta f\left( {\omega '} \right)} \right]}}{{{{\omega '}^2} - {\omega ^2}}}d} \omega ' \textrm{ ,}
\end{align}$\mathcal{P}$ is the Cauchy principal value.

\begin{figure}[htbp]
\centering
\includegraphics[width=0.45\textwidth]{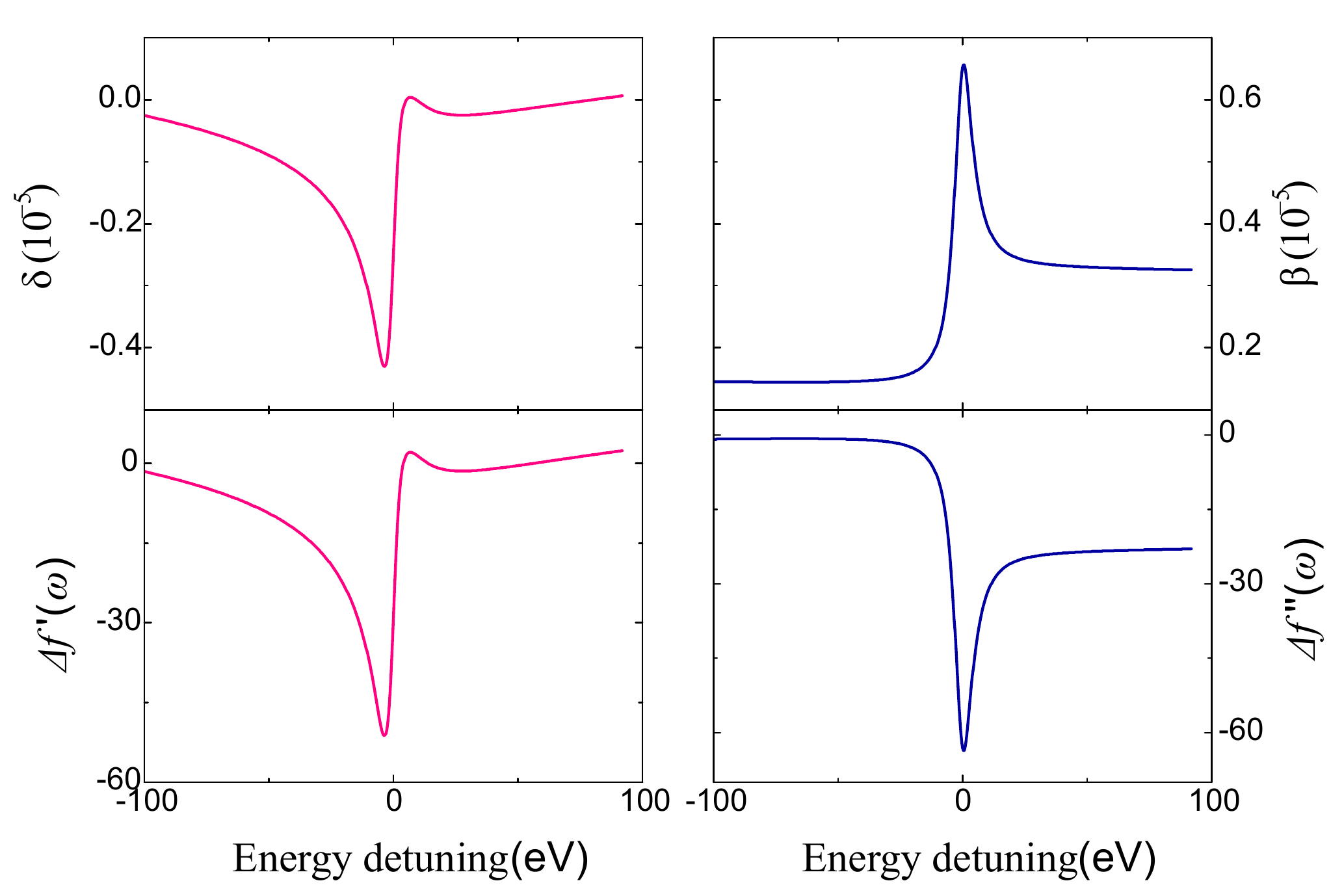}
\renewcommand{\figurename}{Fig.}
\caption{\label{Fig.3}The refractive index deviations $\delta$, $\beta$ and $\Delta f$ of WSi$_2$.}
\end{figure}

\begin{figure}[htbp]
\centering
\includegraphics[width=0.45\textwidth]{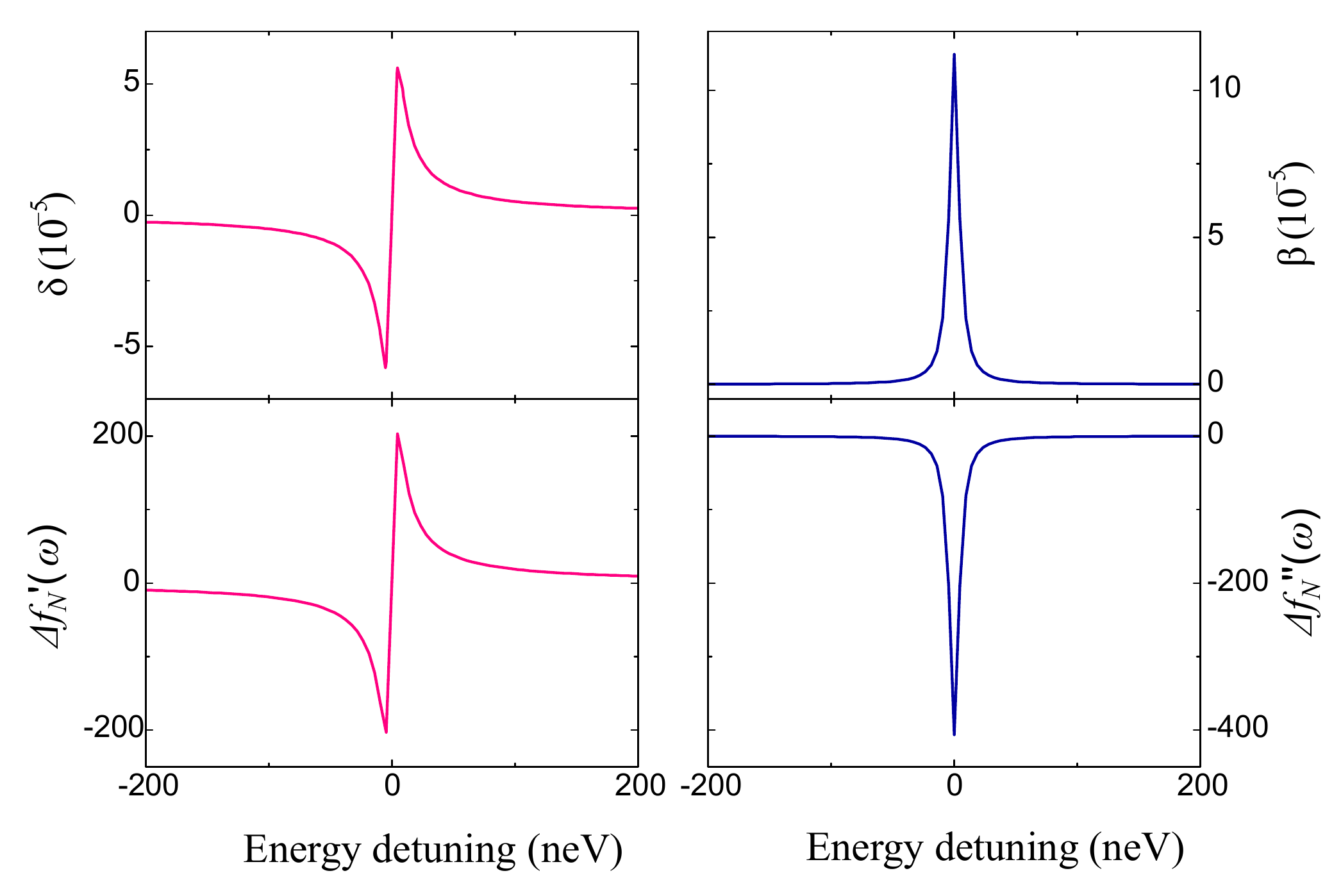}
\renewcommand{\figurename}{Fig.}
\caption{\label{Fig.4}The refractive index deviations $\delta$, $\beta$ and $\Delta f_N$ of $^{57}$Fe~\cite{rohlsberger2004nuclear}.}
\end{figure}

Utilizing Eqs. (9), (12) and (13), and the XAS result of Fig. \ref{Fig.1}(c), the refractive index deviations $\delta$ and $\beta$ and $\Delta f$ of WSi$_2$ are obtained as shown in Fig. \ref{Fig.3}, note that the tradition of $n=1-\delta+i\beta$ is followed. For comparison, the refractive index deviations $\delta$ and $\beta$ and $\Delta f$ of nuclear resonance of $^{57}$Fe are given in Fig. \ref{Fig.4} (The datum are from ref. \cite{rohlsberger2004nuclear}). It can be clearly seen that $\Delta f_N$ is much bigger than $\Delta f$, which will result in a much stronger coupling strength with the cavity mode. This may explain why the cavity effect for electronic resonance system is not as strong as the nuclear one, e.g, almost tens of times decay rate enhancement was observed in the nuclear system~\cite{Rohlsberger1248}, while only about two times decay rate enhancement was achieved~\cite{PhysRevResearch.3.033063} in the electronic system. The nuclear ones show perfect Lorentzian profiles, while the electronic ones are affected from the absorption edge on the right tail because the white line transition overlaps with the electronic continuum. On the other hand, the decay width of the nuclear one is in neV scale which gives hundreds nano seconds lifetime, so probing coherence in time space is possible. While for the electronic one, the linewidth is much bigger, which makes it very hard to observe the cavity effect in the time space. Nevertheless, as learned from above scattering theory, when we deal with the cavity structure in frequency space there is no difference between the nuclear and electronic ones, some typical cavity effects, e.g., CER, CIS and Fano interference, have been also achieved in the electronic systems~\cite{Haber2019,PhysRevResearch.3.033063}.

\subsection{C. Numerical results}
\begin{figure}[htbp]
\centering
\includegraphics[width=0.45\textwidth]{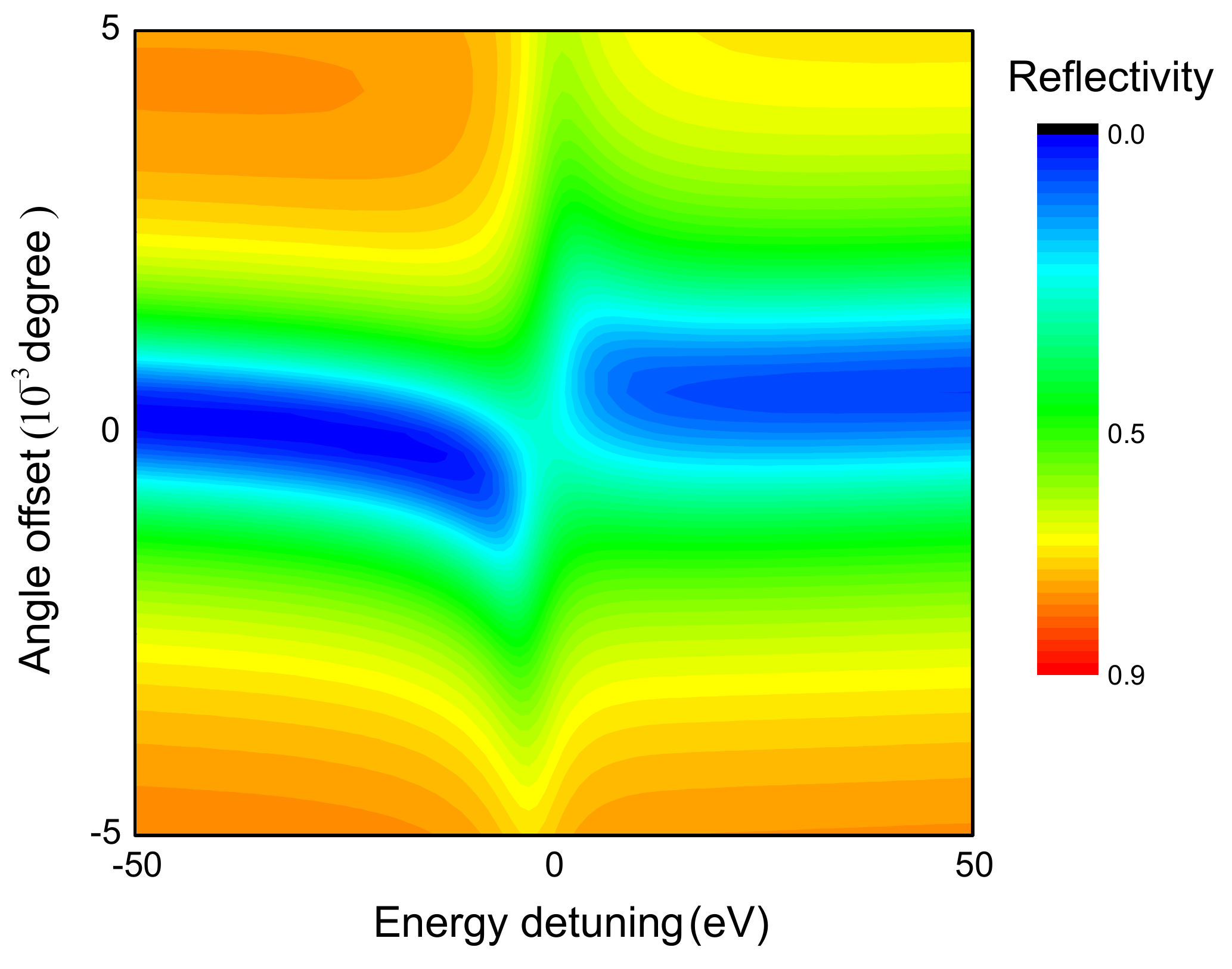}
\renewcommand{\figurename}{Fig.}
\caption{\label{Fig.5}Using the Parratt's formalism, the 2D map of the reflectivity spectra vs. the energy and angle offset is obtained. The cavity structure is shown in Fig. \ref{Fig.1}(a). Angle offset $\Delta \theta$means the deviation between the incident angle $\theta$ and the first cavity mode angle $\theta_1$. A clearly anti-crossing behavior is shown, which is actually from the Fano interference effect, but the Parratt's formalism can not give an interpretation.}
\end{figure}

\begin{figure*}[htbp]
\centering
\includegraphics[width=0.8\textwidth]{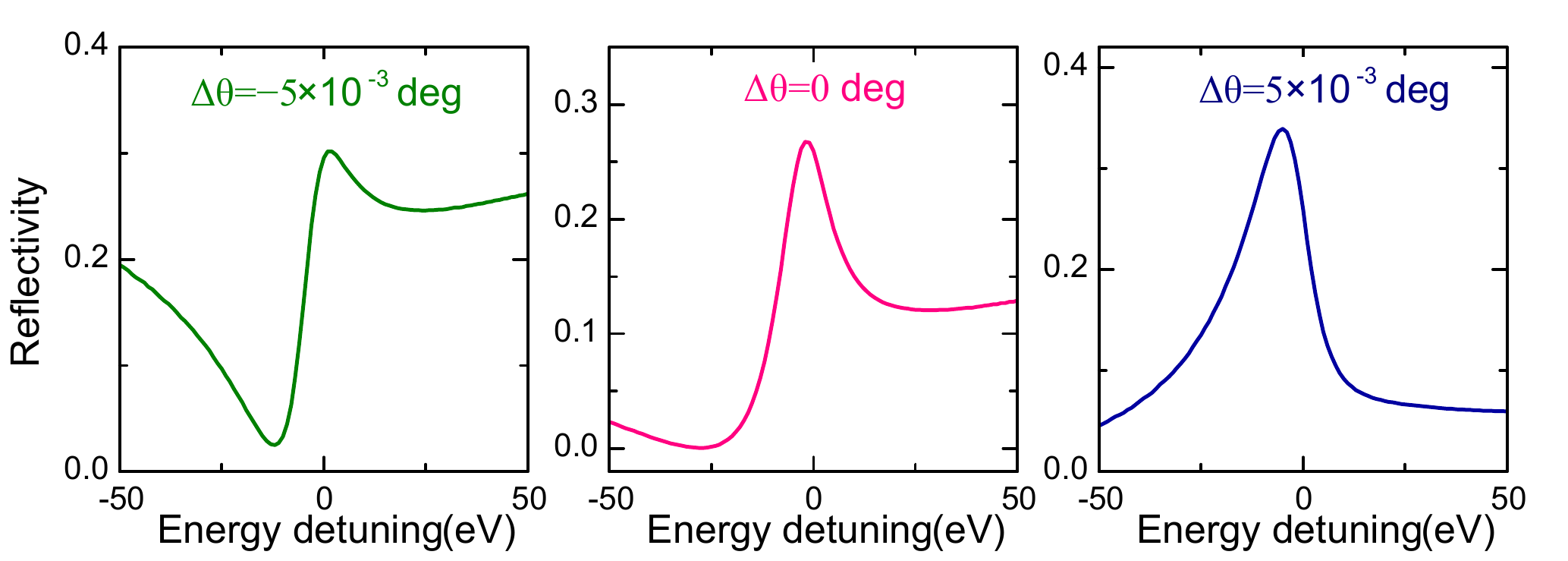}
\renewcommand{\figurename}{Fig.}
\caption{\label{Fig.6}The reflectivity spectra as a function of incident energy in different angle offsets.}
\end{figure*}

Using the refractive indexes of cladding, guiding and atomic layers, we can simulate the reflectivity spectra under any geometries of different incident energies and angles, and the results are depicted in Fig. \ref{Fig.5}. The anti-crossing behavior which have been observed in the experiments~\cite{Haber2019,PhysRevResearch.3.033063} is reproduced, and the peak shift and broadening of the line profile are also shown. This means that the Parratt's method can be well coincident with the experimentally spectral observation, so Fig. \ref{Fig.5} will be used to benchmark the semi-classical and Green's function calculations. The reflectivity spectra in different angle offsets $\Delta \theta$, where $\Delta \theta = \theta - \theta_1$ and $\theta_1$ is the cavity mode angle, are given in Fig. \ref{Fig.6}. When $\Delta \theta=0$ degree, an enhanced resonant elastic scattering is observed. For the ones of $\Delta \theta=\pm 5 \times 10^{-3}$ degree the Fano-like profiles are shown, but we do not know where the two interference channels come from. It can be seen that, the Parratt's formalism can not give a deep interpretation for these observations and the phenomena can not be well understood too, because it only produce an overall reflectance that agrees with the experimental result and leaves out the quantum views.

\begin{figure}[htbp]
\centering
\includegraphics[width=0.45\textwidth]{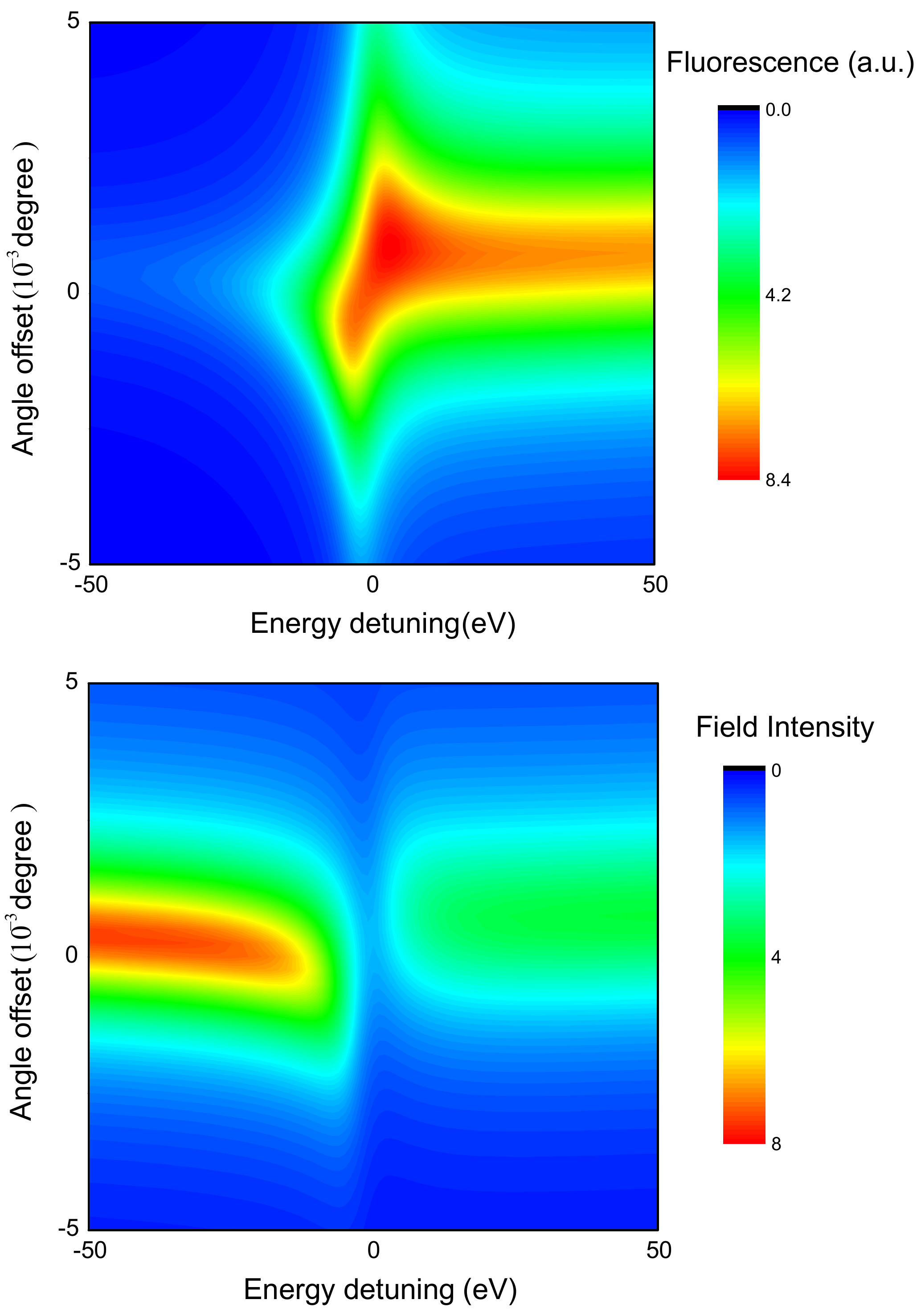}
\renewcommand{\figurename}{Fig.}
\caption{\label{Fig.7}The calculated fluorescence spectra (top panel) and the field intensity map at the atomic layer position (bottom panel) as functions of the incident photon energy and the angle offset. The fluorescence and field intensities show a non-contradictory conformity that is awkward to interpret the line profile broadening and shifting.}
\end{figure}

The field intensities at any frequencies and position inside the cavity could be calculated according to Eqs. (6) and (7), this will be useful to simulate the fluorescence spectrum. Based on the reciprocity theorem, the fluorescence intensity is phenomenologically related to the absorption coefficient $\mu(\omega)$, the field intensities at the atomic position for incident and emitted fluorescence energies~\cite{andre2015kossel,li2014correction}

\begin{equation}
F(\omega,\theta)=c(z_a)\cdot \mu(\omega) \cdot I(\omega,\theta,z_a)\cdot I_f(\omega_f,\theta_f,z_a) \textrm{ ,}
\end{equation}where $c(z_a)$ is a scaling constant. $I(\omega,\theta,z_a)$ is the field intensity at the atomic layer position for incident beam which could be calculated by Eq. (7). $I_f(\omega_f,\theta_f,z_a)$ is the field intensity at the atomic layer position with assuming a virtual source located at the detector position, and $\theta_f$ is the fluorescence emitted angles. In general $\theta_f$ is very large thus $I_f(\omega_f,\theta_f,z_a)$ is almost a unit constant. It can be seen that only $I(\omega,\theta,z_a)$ is adjustable through the incident energy and angle. The simulated fluorescence spectra and the field intensity as functions of incident energy and angle are shown in Fig. \ref{Fig.7}. When the incident angle scans across the first mode angle, a phenomenon of the line profile first broadening to maximum at the mode angle, then narrowing is depicted, also along with the peak shift. Note here that the behaviors of the line profile broadening and peak shift have been observed in the experiments~\cite{Haber2019,PhysRevResearch.3.033063}. We know the standing wave only forms when the incident angle $\theta$ is around the cavity mode angle $\theta_1$, but in Fig. \ref{Fig.7}(b), the field intensity approaches to minimum when the energy closes to the transition energy of the white line while it approaches to maximum when the energy is off-detuned, this makes sense that the line width of fluorescence spectrum is widest when $\theta=\theta_1$, i.e., the spectral behavior of $\mu(\omega)$ is strongly modified by the abrupt field distribution. When the incident angle is detuned from $\theta_1$, the field intensities are flat as a function of incident energy, in this case the fluorescence will be simply connected with the absorption coefficient itself and the fluorescence spectrum shows the natural linewidth. This complex behavior of field redistribution has also been observed in the nuclear system~\cite{huang2017field,PhysRevA.91.063803}, it is a side feature of the cavity effect~\cite{andreeva2008nuclearNIMB} and shows non-contradictory conformity with the evolution of the fluorescence intensity. However, this picture is very awkward, it can not interpret which physical parameters control the width and peak position of the fluorescence spectrum, just gives an overall result in spectral observation.

\section{III. Semi-classical Matrix Method}

\subsection{A. Matrix formalism derivation}

\begin{figure}[htbp]
\centering
\includegraphics[width=0.45\textwidth]{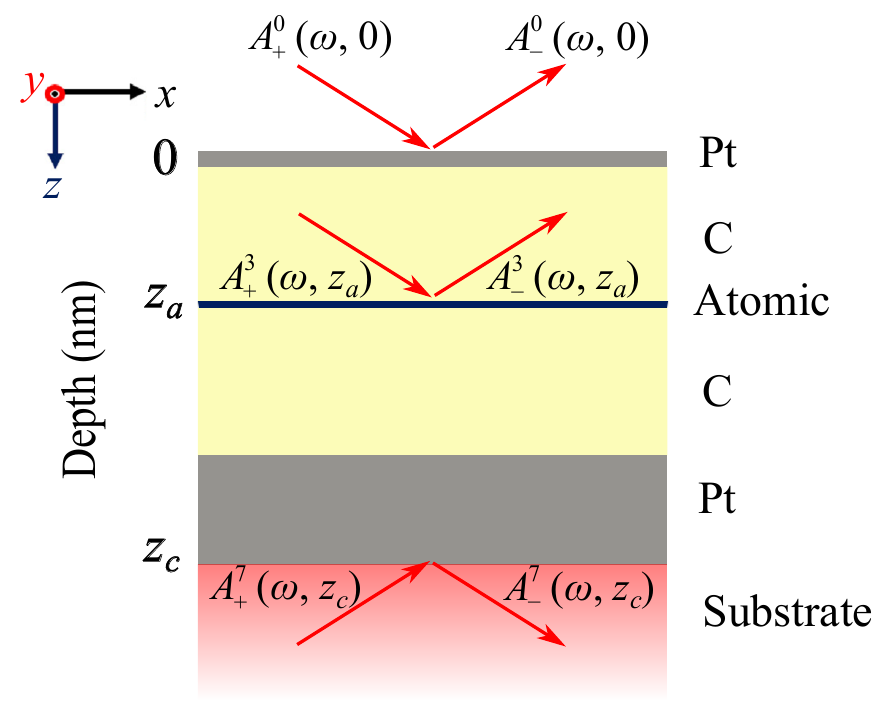}
\renewcommand{\figurename}{Fig.}
\caption{\label{Fig.8}Schematic of the semi-classical matrix model and the cavity geometry in the $xz$ plane.}
\end{figure}

As discussed above, the recursion in Eq. (4) is the core algorithm in Parratt's formalism, and it actually could be equally remoulded as the expression of transfer matrix

\begin{equation}
\left( \begin{array}{l}
A_ + ^j\left( \omega  \right)\\
A_ - ^j\left( \omega  \right)
\end{array} \right) = M_i\left( \begin{array}{l}
A_ + ^i\left( \omega  \right)\\
A_ - ^i\left( \omega  \right)
\end{array} \right) \textrm{ ,}
\end{equation}where

\begin{equation}
M_i = \frac{1}{{{t_{ji}}}}\left( \begin{array}{ccc}
1 &{\rm{   }}{r_{ji}}\\
{r_{ji}}{\rm{  }} &1
\end{array} \right)\left( \begin{array}{ccc}
{\rm{exp}}\left( {i{k_i}{d_i}} \right){\rm{   }}&0\\
0 &{\rm{  exp}}\left( { - i{k_i}{d_i}} \right)
\end{array} \right) \textrm{ .}
\end{equation}Eq. (15) represents that the field amplitude $\left( {A_ + ^j,A_ - ^j} \right)^T$ at layer $j$ is connected with the field amplitude of $\left( {A_ + ^i,A_ - ^i} \right)^T$ at layer $i$ through the prorogating matrix $M_i$. $d_i$ is the thickness of layer $i$. When we deal with the multilayer, the full matrix is obtained by multiplying the propagation matrices of all layers from the surface to the expected depth position. On the other hand, considering~\cite{rohlsberger2004nuclear,PhysRevB.69.235412}

\begin{align}
\left( \begin{array}{ccc}
1{\rm{   }}&{r_{ik}}\\
{r_{ik}}{\rm{  }}&1
\end{array} \right)\left( \begin{array}{ccc}
1{\rm{   }}&{r_{ji}}\\
{r_{ji}}{\rm{  }}&1
\end{array} \right) &=  - \left( \begin{array}{ccc}
1{\rm{   }}&{r_{kj}}\\
{r_{kj}}{\rm{  }}&1
\end{array} \right) \nonumber \\
\frac{{{r_{ji}} + {r_{ki}}}}{{{t_{ik}}{t_{ji}}}} &= \frac{1}{{{t_{ki}}}} \textrm{ ,}
\end{align}$M_i$ could be rewritten as a compact form for diagonalization~\cite{rohlsberger2004nuclear}

\begin{equation}
{M_i} = \textrm{exp}{(i\textbf{F}d_i)} \textrm{ ,}
\end{equation}where $\textbf{F}$ is a $2\times2$ matrix which is directly connected with the atomic scattering length $f$

\begin{equation}
\textbf{F}=\left( \begin{array}{ccc}
f &f\\
 - f &- f
\end{array} \right)+\left( \begin{array}{ccc}
k_i &0\\
0 &-k_i
\end{array} \right) \textrm{ .}
\end{equation}Eq. (19) is the base stone in the semi-classical matrix model. As discussed in Eq. (11), the scattering length $f$ is composed by two parts: the non-resonant scattering $f_T$ and the resonant scattering correction $\Delta f$. Below we will show how the semi-classical matrix model separates the scatterings form the bare cavity ($f_T$ from all layers) and the resonant one ($\Delta f$ from the atomic layer).

The 1D cavity structure is depicted in Fig. 8, it contains overall 7 layers including the air and substrate. Firstly we only consider the bare cavity, i.e, all of the layers and the corresponding refractive indexes are involved, in addition to that of the atomic layer without taking into account the resonant white line transition, so only the background index from $f_T$ and the absorption edge are involved which can be found from the data base~\cite{cxrodatabase}. The field amplitude at the bottom of the cavity $\left( {A_ + ^7,A_ - ^7} \right)^T$ is connected to the one of topmost surface by the complete matrix for the bare cavity $M^{z_c}$

\begin{equation}
\begin{array}{c}
\left( \begin{array}{l}
A_ + ^7\\
A_ - ^7
\end{array} \right) = M^{z_c} \left( \begin{array}{l}
A_ + ^0\\
A_ - ^0
\end{array} \right)\\
\\
M^{z_c} = {M_6} \cdots {M_0} \textrm{ ,}
\end{array}
\end{equation}similar with the Parratt's formalism, using the boundary conditions of $A_ + ^0=A^0$ and $A_ - ^s(\omega)=0$. The reflection coefficient is given,

\begin{equation}
r_0=\frac{A_ - ^0}{A^0}=-\frac{M_{21} ^{z_c}}{M_{22} ^{z_c}} \textrm{ ,}
\end{equation}$M_{nm}, (n, m\leq2)$ is the matrix element. Moreover, considering a generalized field amplitude at any position of $z$ inside the cavity, $M^z$ could be also used to give the normalized field amplitude at position $z$ as similar with Eq. (6)

\begin{align}
a(z) &= \frac{{A_ + (z) + A_ - (z)}}{{{A^0}}} \nonumber \\
&= \left( {M_{11}^z + M_{21}^z} \right) - \left( {M_{12}^z + M_{22}^z} \right)\frac{{M_{21}^{{z_c}}}}{{M_{22}^{{z_c}}}} \textrm{ .}
\end{align}
Here we define two important factors by the matrix elements in $M^z$

\begin{equation}
\begin{array}{lll}
p(z)&=M_{11}^{z}+M_{21}^{z} \\
q(z)&=M_{12}^{z}+M_{22}^{z}
\end{array} \textrm{ ,}
\end{equation}where $p(z)$ is the field amplitude corresponds to the wave from up direction scattered into the up and down directions at the position $z$, and $q(z)$ is the one corresponds to the wave scattered from the down direction into the down and up directions. These two factors describe the field amplitudes that come from the Thomson scattering under the specific cavity structure, moreover it will be also very useful to describe the enhanced resonant x-ray elastic scattering in the reflection direction.

Next we take into account the atomic layer, which is sandwiched at the position of $z_a$, according to Eq. (18) the propagation matrix for the atomic layer only involves the white line resonant transition is written as

\begin{equation}
M^a=\textrm{exp}(i\textbf{F}_a \cdot d) \textrm{ ,}
\end{equation}where $d$ is the thickness of the atomic layer. In usually used geometry, $d$ is very small in few nanometers. In this case, we can apply the ultrathin-film approximation to $M^a$ to expand the exponential function and omit the higher orders

\begin{equation}
\textrm{exp}(i\textbf{F}_a \cdot d)\approx 1+i\textbf{F}_a \cdot d \textrm{ .}
\end{equation}This expansion is the core idea in the semi-classical matrix model. Because the matrix of the bare cavity is already calculated, next we can  obtain the complete matrix for the cavity including the atomic electronic resonance as

\begin{align}
M&=M^{z_c-z_a}(1+i\textbf{F}_a \cdot d)M^{z_a} \nonumber \\
 &=M^{z_c}[1+id(M^{z_a})^{-1} \textbf{F}_a M^{z_a}] \textrm{ ,}
\end{align} with $\textbf{F}_a$ expressed as

\begin{equation}
\textbf{F}_a=\left( \begin{array}{ccc}
\Delta f &\Delta f \\
-\Delta f &-\Delta f
\end{array}\right) \textrm{ ,}
\end{equation}After calculating Eq. (26) with the matrix operations ($\mathbb{I}=M^{-1}M$, $qr_0\approx0$), the reflection coefficient of the whole cavity structure is derived as

\begin{align}
r&=r_0+r_a \nonumber \\
r_a &= \frac{id\Delta f\times |a(z_a)|^2}{1-id\Delta f \times p(z_a)q(z_a)} \textrm{ .}
\end{align}It is clearly seen that the reflectance of the cavity with embedding the atomic layer is from the two pathways: the first one of $r_0$ is from the bare cavity itself where the photon does not interact with the resonant atom, and the second one of $r_a$ is from the enhanced resonant elastic scattering. The thin-film cavity in general has a $Q$ factor of about 100, and the linewidth of the electronic resonance is on the scale of few eV. Therefore, the energy linewidth of $r_0$ is much broader than $r_a$, resulting in the Fano-type interference~\cite{Haber2019,PhysRevResearch.3.033063,vassholz2021observation}. Now we can know that the semi-classical matrix model can give the physical interpretation for the anti-crossing behavior and the Fano profiles. Because the $\Delta f$ only involves the electronic resonance of the white line transition and leave out the edge effect in the present approach, therefore it should be well described by a single Lorentzian function

\begin{equation}
\Delta f=\frac{f_0}{\epsilon+i} \textrm{ ,}
\end{equation}where $\epsilon=\frac{2\delta}{\Gamma}$ is the dimensionless energy, $\Gamma$ is the natural decay width which can be fitted from Fig. \ref{Fig.1}(c), $\Delta=\omega-\omega_0$ is the energy detuning and $\omega_0$ is the transition energy of the white line which can also be fitted from Fig. \ref{Fig.1}(c). $f_0$ is a constant to characterize the resonant scattering length. Inset the Lorentzian function into $r_a$, it is easily to derive that

\begin{equation}
r_a=-\frac{idf_0 \times |a(z_a)|^2}{\Delta +\Delta_c+i(\Gamma+\Gamma_c)/2} \textrm{ ,}
\end{equation}which is still a Lorentzian function, while it contains two additional cavity effects: CER $\Gamma_c$ and CIS $\Delta_c$

\begin{equation}
\begin{array}{lll}
\Delta_c &=df_0 \times \textrm{Im}(\eta) \\
\Gamma_c &=df_0 \times \textrm{Re}(\eta) \\
\eta &=p(z_a)q(z_a) \textrm{ ,}
\end{array}
\end{equation}It can be seen that the strengthes of $\Delta_c$ and $\Gamma_c$ are determined by the imaginary and real parts of the field amplitude $\eta$ of the bare cavity.

\subsection{B. Numerical results}
\begin{figure*}[htbp]
\centering
\includegraphics[width=0.8\textwidth]{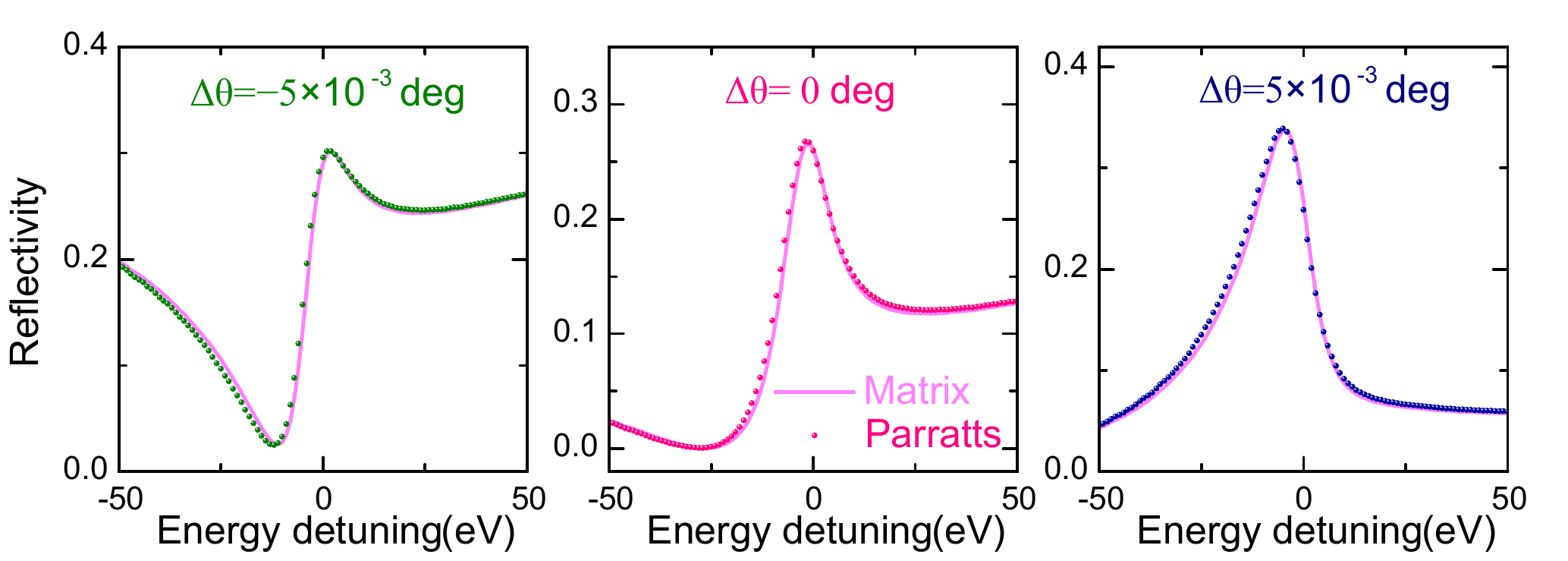}
\renewcommand{\figurename}{Fig.}
\caption{\label{Fig.9}The reflectivity spectra as a function of incident energy in different angle offsets. The curves in dot are calculated by the Parratt's method, and the solid lines represent the calculations of the semi-classical matrix.}
\end{figure*}

\begin{figure}[htbp]
\centering
\includegraphics[width=0.45\textwidth]{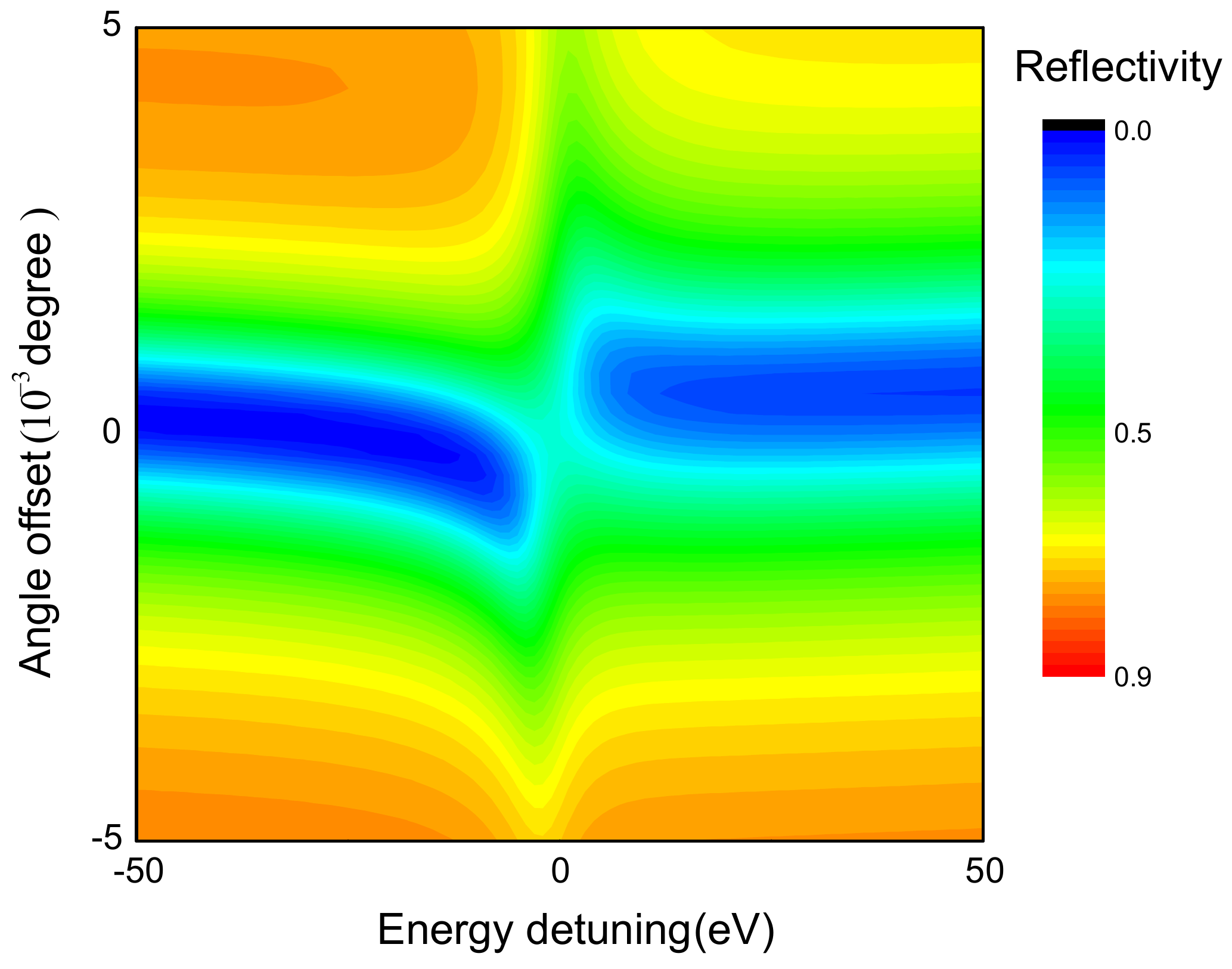}
\renewcommand{\figurename}{Fig.}
\caption{\label{Fig.10}Using the matrix method and the fitted value of $f_0$, the 2D map of the reflectivity spectra vs. the energy and angle offset is obtained. A clearly anti-crossing behavior is shown, which is from the Fano interference effect.}
\end{figure}

We use the Parratt's result as the benchmark to test the validity of the semi-classical Matrix model, the constant $f_0$ is the only one fitting parameter. Using Eq. (28) to fit the reflectivity spectrum at $\Delta \theta=0$ degree, a very good agreement between the Parratt's and matrix methods is shown in the middle panel of Fig. \ref{Fig.9}. The fitted value of $f_0=0.36$ is much smaller than the one of nuclear resonance of $^{57}$Fe $f_0 ^N=2.53$~\cite{Rohlsberger1248}. This value is in line with our expectations, because the larger value of $f_0 ^N$ gives stronger cavity effect, e.g, the peak value of the reflectance for the nuclear resonant scattering can reach to 0.8~\cite{rohlsberger2012electromagnetically}, while for the one of electronic resonance only about 0.25 is obtained when the cavity detuning angle offset is 0 (The middle panel in Fig. \ref{Fig.9}). On the other hand, the reflectivity spectra for other angle offsets are also consistent with the Parratt's results without any further adjustment, and the comparisons are shown in the left and right panels of Fig. \ref{Fig.9} for $\Delta \theta=\pm 5 \times 10^{-3}$ degree. Then the 2D reflectivity map as functions of incident energy and angle offset is obtained as depicted at Fig. \ref{Fig.10}, a very good agreement with Fig. \ref{Fig.5} is observed. And now we can know from the semi-classical model that the anti-crossing behavior is due to the two-pathway Fano interference according to the discussion of Eq. (28).

\begin{figure}[htbp]
\centering
\includegraphics[width=0.45\textwidth]{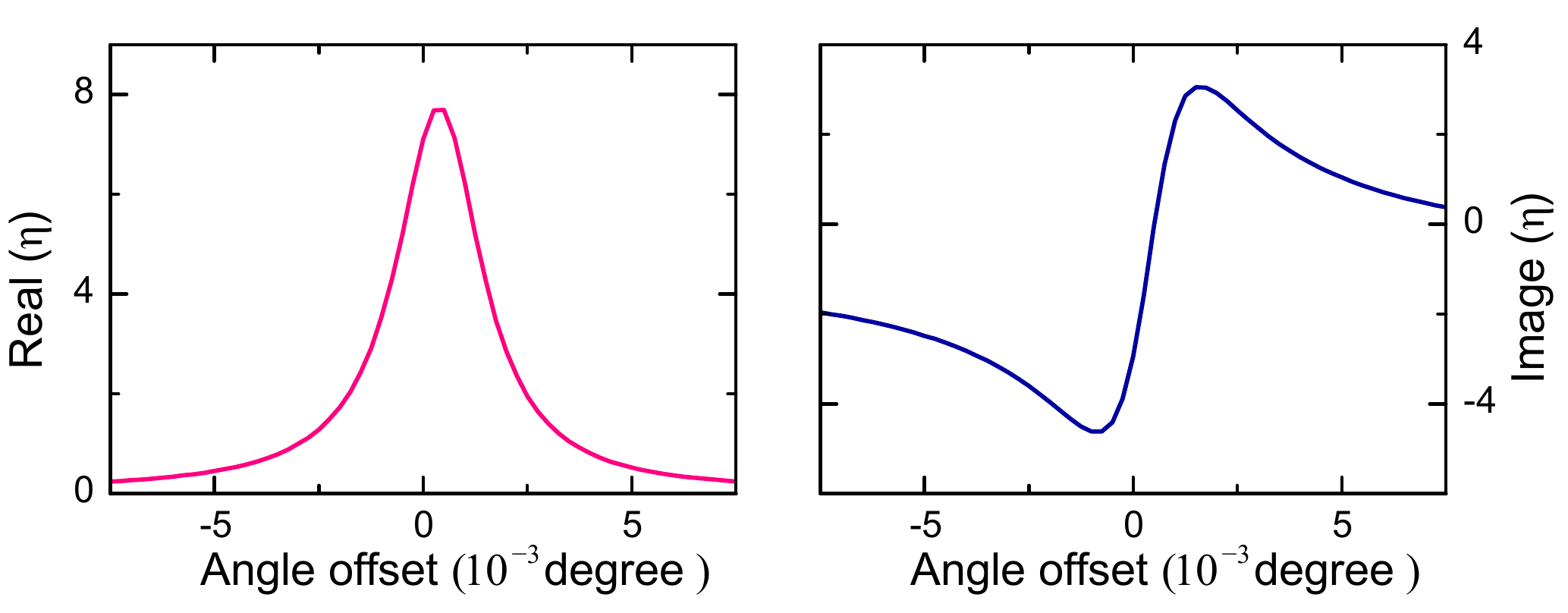}
\renewcommand{\figurename}{Fig.}
\caption{\label{Fig.11}The values of Re($\eta$) and Im($\eta$) as a function of the angle offset, which are calculated by Eqs. (23) and (31).}
\end{figure}

Moreover, the numerical results of the enhanced factor $\eta$ is shown in Fig. \ref{Fig.11}. As depicted at Eq. (31), we know that the emission rate of the electronic resonance is enhanced by a factor of Re($\eta$) and the transition energy is shifted by a factor of Im$(\eta)$. In Fig. \ref{Fig.11}, the value of Re$(\eta)$ shows the maximum when the incident angle $\theta$ equals the cavity mode angle $\theta_1$ ($\Delta \theta=0$ degree). For Im$(\eta)$, its value from 0 firstly decreases to a negative minimum then increases to the positive maximum and finally drops back to 0, which predicts a behavior of firstly negative then positive energy shifts. It can be seen that $\eta$ is the crucial parameter that describes the phenomena of CER and CIS. Actually, it has been successfully used to interpret the linear relation between $\Gamma_c$ and Re($\eta$) in recent experiment~\cite{PhysRevResearch.3.033063}. However, we need to admit here that the physical picture given by the semi-classical matrix method is phenomenological. For example the key factor $\eta$ is a little bit elusive, because the semi-classical model can not connect it with a fundamental physical parameter. Another question is that even though the semi-classical model can give a prediction of CER and CIS from the cavity effect, it is still difficult to derive the effective energy-levels since it does not involve the interaction Hamiltonian. Moreover, the expansibility of the semi-classical model is not very good, we can image the formula simplification from Eq. (28) to Eq. (30) will be tricky when multiple atomic layers are embedded inside the cavity. These questions could only be answered by the quantum model which will be introduced in next section.

\section{IV. Quantum Green's function}
The observed photons scattered by the thin-film planar cavity can be generally attributed to the tailored electromagnetic field by the multilayer stacks of dielectric media. To solve such type problem, a set of electromagnetic Green's functions known as the macroscopic quantum electrodynamics~\cite{scheel2008macroscopic} have been developed for a variety of applications, for example, light-emitting tunnel junction \cite{Ushioda-1984,Ushioda-1986}, Raman scattering \cite{PhysRevA.51.2545}, subradiance and selective radiance of the atmoic arrays \cite{Kimble-2017} and the thin-film cavity with nuclear resonance \cite{PhysRevA.102.033710,PhysRevA.104.033702,PhysRevResearch.2.023396}.
In this section, the modification of the electromagnetic field with electronic resonance in the thin-film planar cavity will be analyzed
from the perspective of quantum Green's function following the framework of Refs. \cite{PhysRevA.102.033710,PhysRevResearch.2.023396}. The analytical expressions of the decay rate, spin-exchange and the inter-layer coupling are derived, which is very useful to give the effective energy-level scheme.

\subsection{A. Model Hamiltonian}
For inner-shell electronic resonance, the white line is generally treated in the dipole approximation~\cite{fink2013resonant,als2011elements,vettier2012resonant,van2016x}. For a normalized and oscillating dipole at $\textbf{r}'$, the Green's function $\rm{\textbf{G}\,(\textbf{r},\rm{\textbf{r}}',\omega)}$ obeys the electromagnetic wave equation given by
\begin{align}
\nabla  \times \nabla  \times \rm{\textbf{G}\,(\textbf{r},\textbf{r}',\omega)} - \varepsilon (\textbf{r},\omega )\frac{{{\omega ^2}}}{{{\textit{c}^2}}}\rm{\textbf{G}\,(r,r',\omega)} = \delta (\textbf{r} - \textbf{r}') \textrm{ .}
\end{align}
Here, $\varepsilon (\textbf{r},\omega )$ is the complex permittivity. Hence, the electric field  ${\rm{\textbf{E}}}(\rm{\textbf{r}},\omega)$ at \textbf{r} with any source current $\textbf{j}(\textbf{r}',\omega)$ can be derived by combining the Green function and given by
\begin{align}
{\rm{\textbf{E}}}(\rm{\textbf{r}},\omega )={\textit{i}\omega \mu_{0}}\int {d\textbf{r}'} \textbf{G}(\textbf{r},\textbf{r}',\omega ) \cdot \textbf{j}(\textbf{r}',\omega ) \textrm{ .}
\end{align}
Especially for the dipole source $\textbf{j}(\textbf{r},\omega)=-i \omega\,\textbf{p}\,\delta(\textbf{r}-\textbf{r}')$ such as the resonant x-ray elastic scattering of white line, the electric field ${\rm{\textbf{E}}}(\rm{\textbf{r}},\omega)$ has the simplest form given by
\begin{align}
\textbf{E}(\textbf{r},\omega ) = {{\mu_{0}\omega ^2}}\textbf{G}(\textbf{r},\textbf{r}',\omega ) \cdot \textbf{p} \textrm{ .}
\end{align}
Here, $\mu_{0}$ is the vacuum permeability. $\textbf{p}$ is the momentum operator.

To this end, for the elastic x-ray scattering of atom as mentioned in Sec. II B, the scattered electric field of the superposition of the Thomson scattering and resonant scattering can be separated into two parts, which is similar with Eq. (11), and given by
\begin{align}
{{\textbf{E}} }(\textbf{r},\omega ) = {\textbf{E}}_{\rm{T}} (\textbf{r},\omega ) + {\textbf{E}}_{\rm{r}} (\textbf{r},\omega ) \textrm{ ,}
\end{align}
where the subscripts `T' and `r' represent Thomson scattering and resonant elastic scattering respectively. For the resonant scattering process, it has the Lorentz shape as discussed in Eq. (29) and can be regarded as the dipole source under dipole approximation. Hence, the scattered electric field can be simplified with the Green's function
\begin{align}
{{\textbf{E}} }(\textbf{r},\omega ) = {\textbf{E}}_{\rm{T}} (\textbf{r},\omega ) + {{\mu_{0}\omega ^2}}\sum\nolimits_{i = 1}^N {{\textbf{p}_i} \cdot \textbf{G}(\textbf{r},\textbf{r}_{i}',\omega )} \textrm{ ,}
\end{align}
The summarization runs over all scatterers at different source position $\textbf{r}_{i}'$. In order to introduce the quantum Green function, the classical field should be translated to the quantum field firstly. Due to the consistency of the Maxwell's equations for field propagation \cite{Kimble-2017}, the dipole moment and classical field only need to do the substitution with
\begin{align}
\textbf{p} &\to \hat {\textbf{p}} \to {\textbf{d}^ * }{\hat {\sigma} ^ + } + \textbf{d}{\hat {\sigma} ^ - } \textrm{ ,}\\
\textbf{E} &\to \hat {\textbf{E}} \to {\hat {\textbf{E}}^ + } + {\hat {\textbf{E}}^ - } \textrm{ ,}
\end{align}
where $\sigma^+ = \left| e \right\rangle \left\langle g \right|$ is the atomic raising operator and  $\sigma^-$ is its conjugate operator. $\textbf{d} = \left\langle g \right|\hat {\textbf{p}} \left| e \right\rangle $ is the dipole matrix element~\cite{Kimble-2017}. The superscript `+' and `-' of the ${\hat {\textbf{E}}}$ present the positive and negative frequency component of the field operators respectively, which are relevant to the bosonic creation operator  $\hat{\rm{\textbf{f}}}(\textbf{r},\omega )$ such as
\begin{align}
\hat {\textbf{E}}^+ (\textbf{r},\omega ) = i{\mu _0}{\omega ^2}\sqrt {\frac{{\hbar {\varepsilon _0}}}{\pi }} \int {\sqrt {{\mathop{\rm Im}\nolimits} [\varepsilon (\textbf{r}',\omega )]} } \textbf{G}(\textbf{r},\textbf{r}',\omega )\hat {\textbf{f}}(\textbf{r}',\omega )d\textbf{r}' \textrm{ .}
\end{align}

Considering the $N$ two-level atoms inside the medium that interact with the electromagnetic field, the general Hamiltonian for such coupling system under rotating-wave and dipole approximation reads~\cite{Dung-2002,Hughes-2009}
\begin{align}
 \hat H = &\int {d\textbf{r}}\int_{0}^{\infty } {\hbar \omega \; {{\hat {\rm{\textbf{f}}}}^\dag }(\textbf{r},\omega )\hat {\rm{\textbf{f}}}(\textbf{r},\omega ) \; d\omega } + \sum\limits_{i = 1}^N {\frac{1}{2}} \hbar {\omega _0} \,\sigma _i^z \nonumber\\
  & -\sum\limits_{i = 1}^N  \int_0^\infty  {{\hat{\textbf{p}}_i}} \cdot \hat {\rm{\textbf{E}}}({\textbf{r}_i},\omega)\; d\omega \textrm{ ,}
\end{align}
where $\omega_0$ is the resonant energy and $\sigma^z = \left| e \right\rangle \left\langle e \right| - \left| g \right\rangle \left\langle g \right| $.
Traced out the freedom of the photonic degree within the Born-Markov approximation,  the effective Hamiltonian in the rotating wave approximation and the Lindbland operator read \cite{Kimble-2017}
\begin{align}
{\hat {H}_{\rm{eff}}} = \sum\limits_{i = 1}^N {\frac{{\hbar {\Delta}}}{2} \hat{\sigma} _i^z}  - \hbar \sum\limits_{i,j = 1}^N {{J_{ij}} \hat{\sigma} _i^ + } \hat{\sigma} _j^ - - \sum\limits_{i = 1}^N {[{d^*} \cdot \textbf{E}_{\rm{T}}^ + ({\textbf{r}_i})} \hat{\sigma} _i^ +  + \rm{H.c.}] \textrm{ ,}
\end{align}
and
\begin{align}
L[\hat {\rho} ] = \sum\limits_{i.j = 1}^N {\frac{{{\Gamma _{ij}}}}{2}(2\hat{\sigma} _i^ - \hat{\rho} \hat{\sigma} _j^ +  - \{ \hat{\sigma} _j^ + \hat{\sigma} _i^ - ,\hat{\rho} \} )}.
\end{align}
Here, $\Delta = \omega - \omega_0$ is the detuning between the cavity field and the resonant transition. The spin-exchange $J_{ij}$ and  decay rates $\Gamma_{ij}$ are given by
\begin{align}
{J_{ij}} &= \frac{{{\mu _0}{\omega ^2}}}{\hbar }{\textbf{d}^*} \cdot {\mathop{\rm Re}\nolimits} [\textbf{G}(\textbf{r},\textbf{r}',\omega )] \cdot \textbf{d} \textrm{ ,}\\
{\Gamma_{ij}} &= \frac{{{2\mu _0}{\omega ^2}}}{\hbar }{\textbf{d}^*} \cdot {\mathop{\rm Im}\nolimits} [\textbf{G}(\textbf{r},\textbf{r}',\omega )] \cdot \textbf{d} \textrm{ .}
\end{align} According to the effective Hamiltonian and the Lindbland operator, it is clear that the spin-exchange and decay rate characterize the energy shift and the decay rate enhancement. These two parameters are controlled by the Green's function which can be modified by the cavity. In section B, we will apply the Green's function to the specific thin-film cavity structure.

\subsection{B. Green's function for thin-film planar cavity}
In the general case, the thin-film cavity works at very small incident angle, i.e., the grazing incidence case, the electromagnetic field will be tailored by the thin-film planar cavity along the $z$ axis as shown in Fig. \ref{Fig.8}. As a consequence, the Green's function is such a system that can be simplified by considering the quasi-1D structure
\begin{align}
\textbf{G}({\textbf{r}_i},{\textbf{r}_j},\omega ) = \frac{1}{{{{(2\pi )}^2}}}\int {{\textbf{G}}({z_i},{z_j},\omega ,{\textbf{k}_{xy}}){e^{i{\textbf{k}_{xy}} \cdot ({\textbf{r}_{xy,i}} - {\textbf{r}_{xy,j}})}}} {d^2}{\textbf{k}_{xy}} \textrm{ ,}
\end{align}
For the weak polarization dependence at a small incidence angle where the cavity modes are driven, the Green's function ${\textbf{G}}({z_i},{z_j},\omega ,{\textbf{k}_{xy}})$ can be further derived from the matrix formalism as mentioned in Sec. III~\cite{PhysRevA.104.033702}
\begin{align}
{\textbf{G}}({z_i},{z_j},\omega ,{\textbf{k}_{xy}}) & \approx \frac{i}{{2{k_z}}}[p({z_i})q({z_j})\Theta ({z_i} - {z_j}) \nonumber\\
& + p({z_j})q({z_i})\Theta ({z_j} - {z_i})] \textrm{ ,}
\end{align}
$\Theta$ is the step function. As depicted in Eqs. (23) and (46), the Green's function for such 1D structure is calculated from the bare cavity. We know that the electron density of the atomic layer is much larger than the guiding layers, i.e., the Thomson scattering and the absorption strength of the atomic layer are much stronger than the ones of guiding layers, so the atomic layer will disturb the accuracy of the Green's function more or less. When the atomic layer is thin enough, all atoms could be approximately regarded as locating at same position of $z$. But if the atomic layer is relative thicker, we need to slice the atomic layer into a set of sublayers. Note here the slice approach will not change the Green's function framework, because when we deal with a medium containing $N$ atoms, summarizations of the spin-exchange and decay rate are considered for all atoms. Therefore, the atomic layer could be arbitrarily sliced into the sublayers $N_l$, so that the loweing operator of the resonant atoms of the atomic layer reads
\begin{align}
\hat {\sigma} _{l}^ - ({\textbf{k}_{xy}}) = \frac{1}{{\sqrt {{N_l}} }}\sum\limits_{i = 1}^{{N_l}} {\hat {\sigma} _i^ - {e^{ - i{\textbf{k}_{xy}} \cdot {\textbf{r}_{xy,i}}}}} \textrm{ ,}
\end{align}
Hence, the effective Hamiltonian considering different multiple sublayers with the interlayer coupling is given by
\begin{align}
{\hat {H}_{\rm{eff}}} &= \sum\limits_{l}{\frac{{\hbar {\Delta}}}{2} \hat{\sigma} _l^z(\textbf{k}_{xy})}  - \hbar \sum\limits_{ll'} {{J_{ll'}(\textbf{k}_{xy})} \hat{\sigma} _l^ +(\textbf{k}_{xy}) } \hat{\sigma} _{l'}^ -(\textbf{k}_{xy}) \nonumber\\
 &- \hbar\sum\limits_{l} {[ \Omega_{l}\hat{\sigma} _l^ +(\textbf{k}_{xy})  + \rm{H.c.}]} \textrm{ ,}
\end{align}
with the Lindbland term to character the incoherent processes~\cite{PhysRevResearch.2.023396}
\begin{align}
L[\hat {\rho}(\textbf{k}_{xy}) ]&= \sum\limits_{ll'} {\frac{{{\Gamma _{ll'}}}}{2}(2\hat{\sigma} _l^-(\textbf{k}_{xy}) \hat{\rho} \hat{\sigma} _{l'}^ +(\textbf{k}_{xy}) - \{ \hat{\sigma} _{l'}^ +(\textbf{k}_{xy}) \hat{\sigma} _l^ -(\textbf{k}_{xy}) ,\hat{\rho} \} )} \nonumber\\
&+L_{\rm{SE}}[\hat {\rho}] \textrm{ ,}
\end{align}
where
\begin{align}
{J_{ll'}} &= \frac{\sqrt{N_{l}N_{l'}}}{A}\frac{{{\mu _0}{\omega ^2}}}{\hbar }{\textbf{d}^*} \cdot {\mathop{\rm Re}\nolimits} [\textbf{G}(z,z',\omega,\textbf{k}_{xy} )] \cdot \textbf{d} \textrm{ ,}\\
{\Gamma_{ll'}} &= \frac{\sqrt{N_{l}N_{l'}}}{A}\frac{{{2\mu _0}{\omega ^2}}}{\hbar }{\textbf{d}^*} \cdot {\mathop{\rm Im}\nolimits} [\textbf{G}(z,z',\omega,\textbf{k}_{xy} )] \cdot \textbf{d} \textrm{ .}
\end{align}
So not only the inter-atom but also the inter-layer couplings are given, and the summarization of all atoms give the overall CER $\Gamma_c$ and CIS $\Delta_c$. $\Omega_{l}=\frac{\sqrt{N_{l}}}{\hbar}{\textbf{d}_{l}^*} \cdot \textbf{E}_{\rm{T}}^ + ({\textbf{z}_l,\textbf{k}_{xy}})$ is the Rabi frequency of the inner-shell resonant transition for sublayer $l$.  $A$ was defined as the parallel quantization area~\cite{PhysRevResearch.2.023396} to translate the Green's function in three dimensional to the quasi one dimensional in the case of grazing incidence, so that the area number density is relevant to the inter-layer coupling and decay rate. According to Ref. \cite{PhysRevResearch.2.023396}, $N_l/A$ can be expressed as

\begin{equation}
\frac{N_l}{A}=\rho_a \cdot d_l \textrm{ ,}
\end{equation}where $d_l$ is the thickness of the sublayer $l$ and $\rho_a$ is the atomic number density.

In the Heisenberg representation, Eq. (47) can be solved in the frequency space, then the motion equation is written as
\begin{align}
{\dot{\hat {\sigma}} _l^-}(\textbf{k}_{xy}) = i(\Delta  + i\frac{{{\gamma _0}}}{2}){\hat{\sigma} _l^-}(\textbf{k}_{xy}) + i{\Omega _l} + i\sum\limits_{l'} {G_{ll'}} {\hat{\sigma} _l^-}(\textbf{k}_{xy}) \textrm{ ,}
\end{align}
and
\begin{align}
{\sigma _l}^ - (\textbf{k}_{xy}) =  - \sum\limits_{l'} {{\mathbb {M}}_{ll'}^{ - 1}} {\Omega _{l'}} \textrm{ ,}
\end{align}
where
\begin{align}
{{\mathbb{M}}_{ll'}} = (\Delta  + i\frac{{{\gamma _0}}}{2}){\mathbb{I}} + G_{ll'} \textrm{ ,}
\end{align}where $\mathbb{I}$ is the unity matrix, and $G$ is the $N_l$ dimension matrix that contains the elements $G_{ll'}$

\begin{align}
G_{ll'}&=J_{ll'}+i\Gamma_{ll'}/2 \nonumber \\
&=\frac{\sqrt{N_{l}N_{l'}}}{A}\frac{{{\mu _0}{\omega ^2}}}{\hbar }{\textbf{d}^*} \cdot \textbf{G}(z,z',\omega,\textbf{k}_{xy}) \cdot \textbf{d} \textrm{ .}
\end{align}

\begin{figure}[htbp]
\centering
\includegraphics[width=0.5\textwidth]{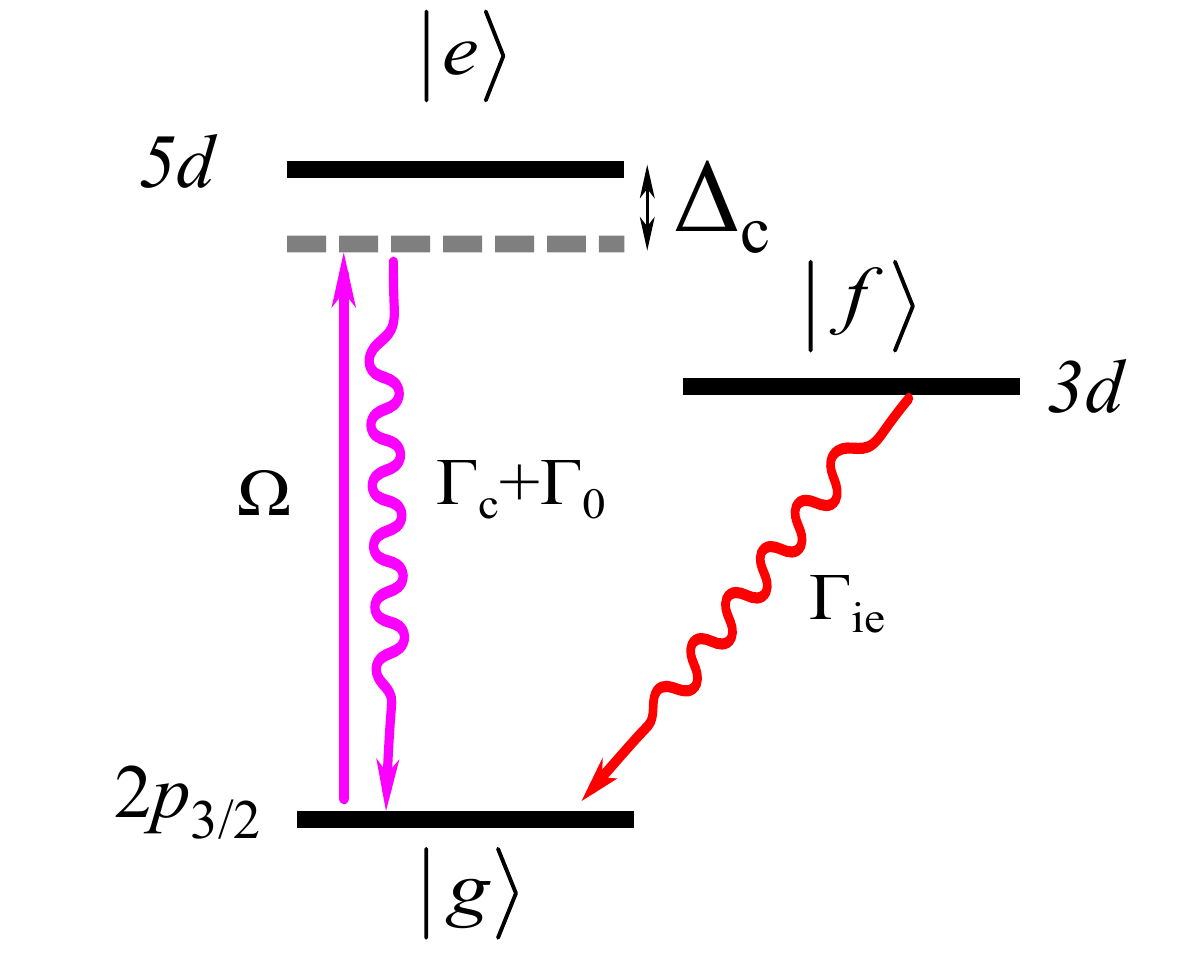}
\renewcommand{\figurename}{Fig.}
\caption{\label{Fig.12}The effective energy-levels around the tungsten L$_{\textrm{III}}$ edge of WSi$_2$ inside the thin-film cavity system. The driving field is shown by the solid arrow, and the decays are represented by the wavy arrows. $\Gamma_0$ is the natural decay rate of the resonant scattering channel, and $\Gamma_{\textrm{ie}}$ is the decay rate of the inelastic relaxations, here the L$\alpha$ fluorescence processes are depicted as example. The summarization of $\Gamma_0$ and $\Gamma_{\textrm{ie}}$ could give the natural decay width $\Gamma$, and the cavity enhanced decay rate is $\Gamma_c$. A simultaneous effect of cavity induced energy shift is also shown, the shift value is labeled by $\Delta_c$.}
\end{figure}

According to the framework of the quantum Green's function, it is clear to see that the CER and CIS are connected to the physical parameters of decay rate and spin-exchange, which are enhanced by the imagery and real parts of the Green's function inside the thin-film cavity respectively. Based on the system Hamiltonian of Eq. (48), the effective energy level is given in Fig. \ref{Fig.12}. The atomic ensemble is driven by the field $\Omega$ that is modified by the Thomson scattering of the specific bare cavity structure, from the ground state $|g\rangle$ to the excited state $|e\rangle$. Due to the cavity effect, the energy and the lifetime of $|e\rangle$ are modified simultaneously, resulting in that $|e\rangle$ is different from the one of atom in free space. For example, the CIS of about 3.0 eV and CER of about 6 eV were realized in Ref. \cite{Haber2019}. Moreover, $|e\rangle$ is actually the intermediated core-hole state, and the particularity of the core vacancy is the extensive relaxation pathways. The other incoherent decay channels such as Auger and inelastic fluorescence processes compete with the second step of the resonant scattering process $|e\rangle\rightarrow |g\rangle$, and the core hole lifetime is determined by the total decay rate of all channels. Normally when the atom is in the free space, the Auger processes dominate the decay routes of the K shell for low-Z atoms~\cite{Auger1925} and L/M shells for higher-Z atoms~\cite{Krause1979}. However, when the resonant scattering process is enhanced through the cavity effect, the decay of the core-hole state will be accelerated which has been reported recently~\cite{PhysRevResearch.3.033063}. Moreover, a maximum value of $\Gamma_c=$ 5.0 eV was realized~\cite{PhysRevResearch.3.033063} which even breaks the limitation of natural decay rate.

\subsection{C. Input-output formalism}

\begin{figure*}[htbp]
\centering
\includegraphics[width=0.8\textwidth]{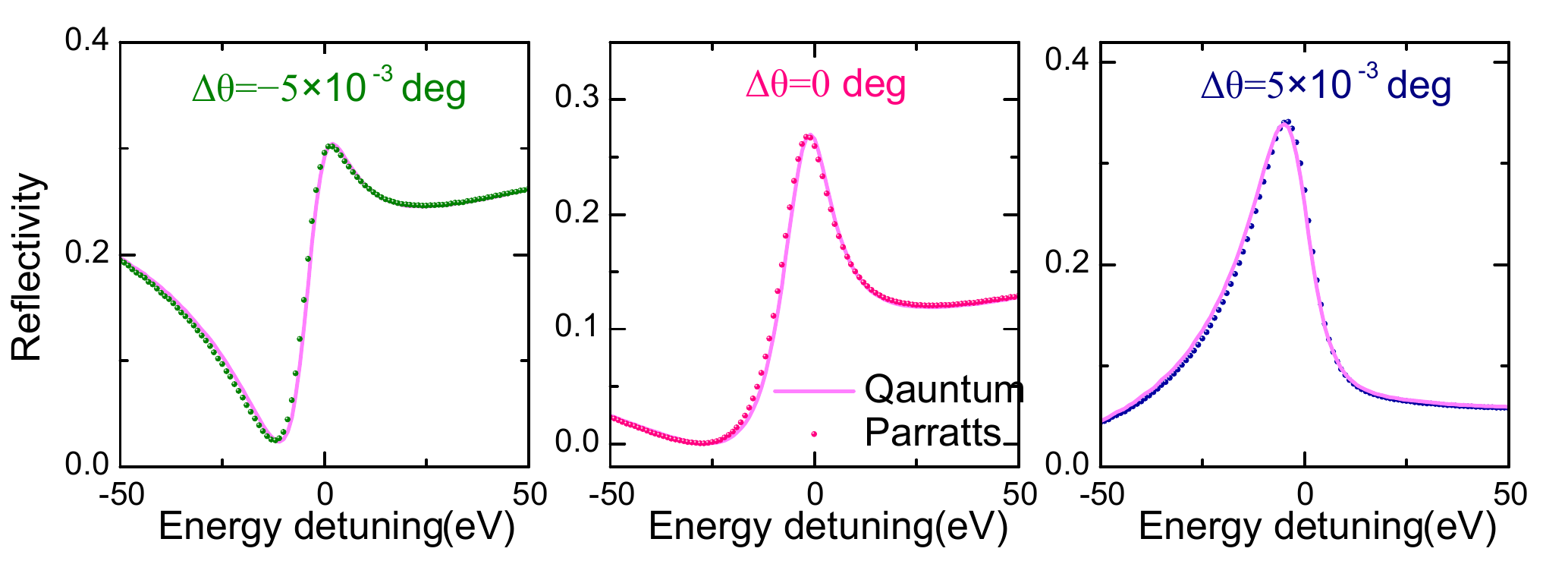}
\renewcommand{\figurename}{Fig.}
\caption{\label{Fig.13}The reflectivity spectra as a function of incident energy in different angle offsets. The curves in dot is calculated by the Parratt's method, and the solid line represent the calculations of the quantum Green's function model. Good agreements between the quantum Green's function model and Parratt's method are observed.}
\end{figure*}

As discussed at the above two sections, a general spectrum observed in experiment is the energy and angle dependent reflectance. In the linear regime, this observation is readily to be got through the input-output formalism, wherein the reflection is calculated by the output field at the surface of cavity and the input field~\cite{PhysRevA.104.033702}

\begin{equation}
R = \frac{{\left\langle {{\hat a_{\textrm{out}}}} \right\rangle }}{{{a_{\textrm{in}}}}} \textrm{ ,}
\end{equation} where

\begin{equation}
{\hat{a}_{\textrm{out}}} = \hat {\textbf{E}}_{\textrm{out}}^ + ({z_{s}}){e^{i{\textbf{k}_{xy}} \cdot {\textbf{r}_{xy}}}} \textrm{ ,}
\end{equation}
is the output operator, and $a_{\textrm{in}}$ is the input field. According to Eq. (35), $\hat {\textbf{E}}_{\textrm{out}}^ + (z)$ contains two components
\begin{align}
\hat {\textbf{E}}_{\textrm{out}}^ + (z) &= \hat {\textbf{E}}_{\rm{T}}^ + (z) \nonumber \\
&+ \frac{{{\mu _0}{\omega ^2}}}{A}\sum\limits_l {\sqrt {{N_l}} } \textbf{G}(z,{z_l},\omega ,{\textbf{k}_{xy}}) \cdot \textbf{d}\sigma _l^ - ({\textbf{k}_{xy}}) \textrm{ ,}
\end{align}where $\hat {\textbf{E}}_{\rm{T}}^ + (z)$ is the Thomson scattering that only from the bare cavity, and the second term is the enhanced resonant scattering from the atomic layer. Under the framework of the quantum optical model, it is easy to understand that the observed reflection of the cavity is from the two separated pathways: bare cavity and the resonant scattering.

\subsection{D. Numerical results}

\begin{figure}[htbp]
\centering
\includegraphics[width=0.45\textwidth]{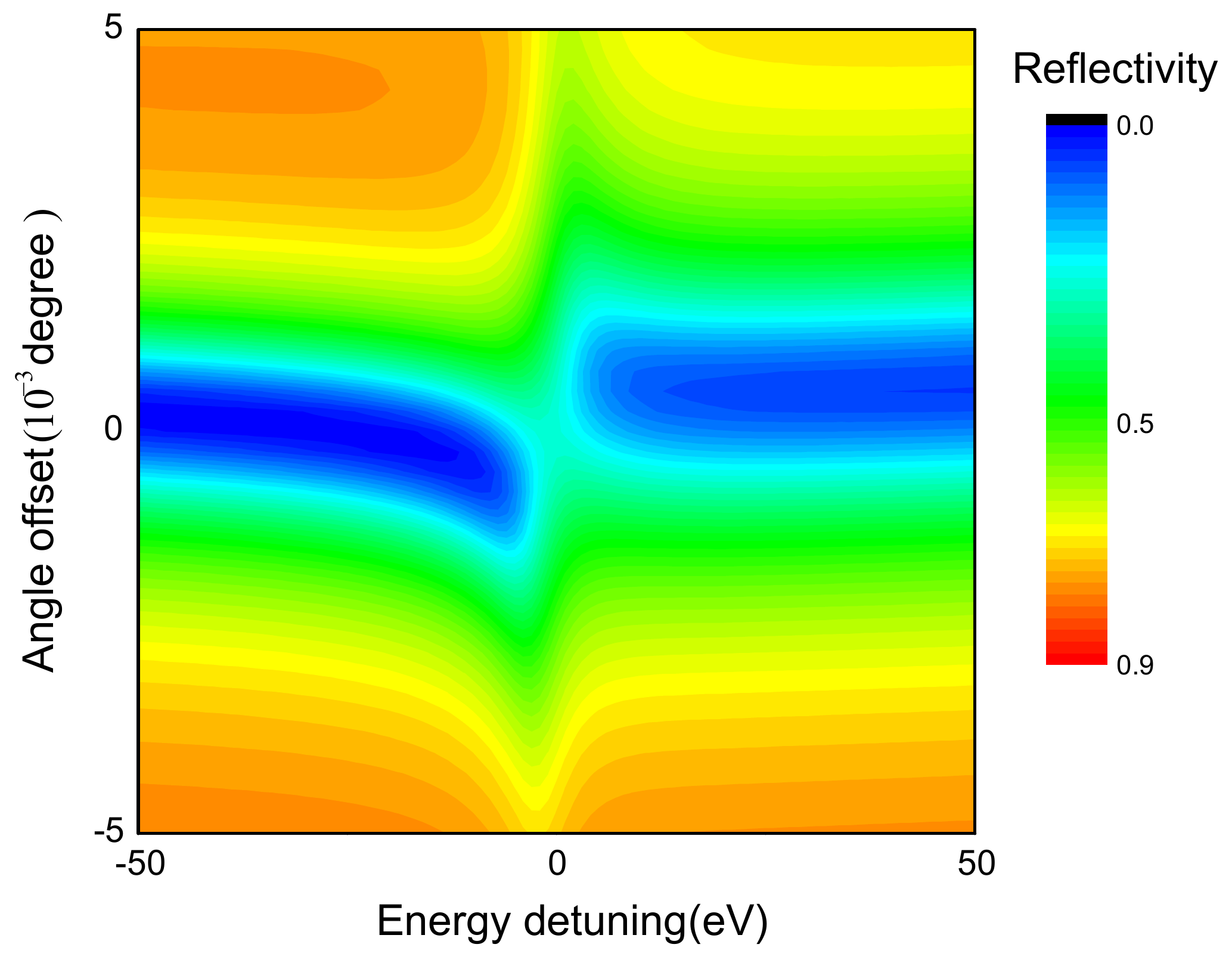}
\renewcommand{\figurename}{Fig.}
\caption{\label{Fig.14}Using the quantum Green's function model and the fitted value of the modular square of dipole moment $|d|^2$, the 2D map of the reflectivity spectra vs. the energy and angle offset is obtained. A clearly anti-crossing behavior is shown, which is from the Fano interference. The results agree well with the ones of Parrtt's and matrix methods.}
\end{figure}

To test the validity of the quantum Green's function model, the Parratt's results are also used as the benchmark. With the assistance of the input-output formalism, there is only one factor need be scaled, the modular square of dipole moment $|d|^2$. Firstly, the reflectivity spectrum of $\Delta \theta=0$ degree is used to obtain the value of $|d|^2$, and a very good agreement between the Green's function and the Parratt's results is shown in the middle panel of Fig. \ref{Fig.13}. The fitted value of $|d|^2$ is $3.3\times 10^{-7}$ (in unit $\Gamma$), which is much smaller than the nuclear dipole moment $|d_N|^2=1.8 \times 10^{-6}$~\cite{PhysRevResearch.2.023396}. The coupling strength between the atom and cavity mode is linearly related to the strength of the dipole moment~\cite{fox2006quantum}, this explains why the electronic resonance systems has the relatively weaker cavity effect. As shown in the left and right panels of Fig. \ref{Fig.13}, without any further adjustment, the reflectivity spectra of other angle offsets also show very good agreements with the Parratt's results, and the Fano-like profiles are observed. Then the 2D reflectivity map as functions of incident energy and angle offset is obtained as depicted at Fig. \ref{Fig.14}, which agrees well with Fig. \ref{Fig.5} of Parratt's model and Fig. \ref{Fig.10} of the semi-classical matrix model. And the Green's function model interprets the anti-crossing behavior very clearly, which is from the two-pathway Fano interference effect.

Under the linear regime, we can see that the quantum Green's function model shows an intrinsic self-consistence with the Kramers-Heisenberg formula of Eq. (10), and it could be intuitively understood as a dedicated and succinct solution of Kramers-Heisenberg scattering theory under the specific cavity-based structural media. We emphasise that the Green's function model provide a framework to separate the Thomson scattering (and background absorption) and the resonant elastic scattering of the atoms, and to individually discuss the atomic response, which is the crucial concept to understand the cavity effect and derive the system Hamiltonian and the effective energy-level. We also need to note here that the Thomson scattering and background absorption strength of the atomic layer are not very small, or said that the resonant scattering is not very strong compared with the background~\cite{als2011elements,attwood2000soft,cxrodatabase,gel1999resonant}. When the thickness of the atomic layer increases, even though the atom number becomes lager, the strength of Green's function of the bare cavity should be unfortunately weakened. This disturbing influence may make the collective effect inapparent in the electronic system which is quite different from the nuclear one wherein the collective strength can be easily controlled by the nuclear layer thickness~\cite{Rohlsberger1248} or the isotope abundance~\cite{li2022probing}.

Different from the nuclear resonance system~\cite{PhysRevResearch.2.023396} wherein the dipole moment is an already known parameter which is connected to the constants of internal conversion factor, spins and the natural decay rate, the quantum Green's function model for the electronic resonance systems reported here is not \emph{ab initial} because the dipole moments of the inner shell white line transitions are usually unknown. On the other hand, for the inner shell transition in the hard x-ray regime, the white line transition around the main edge always overlaps with the absorption edge which also involves the dipole transition~\cite{PhysRevA.22.1104} and even can be approximately modelled by a series of dipole oscillators with different weights~\cite{als2011elements}. Nevertheless, the quantum Green's function model provide a \emph{ab initial} framework. Combining with the accurate quantum chemistry calculations for the inner-shell transitions~\cite{de2008core,de2001high,debeer2008prediction,levine2009quantum}, the quantum Green's function model is possible to be complete \emph{ab initial}.

The Green's function model indicates that the cavity effect acts on the modification of the intermediated core-hole state, wherein the energy and core-hole lifetime are manipulated which are different from the natural ones. Moreover, one of the relaxation pathways of the core-hole state, i.e, the resonant elastic scattering process is enhanced. Actually, the core-hole state is the basic concept in a variety of modern x-ray spectroscopy techniques~\cite{de2008core,schulke2007electron}, e.g., its elastic relaxation pathway gives the resonant elastic x-ray scattering (REXS) spectroscopy~\cite{fink2013resonant,vettier2012resonant,abbamonte2002structural}, its inelastic fluorescence pathways give the resonant x-ray emission spectroscopy (RXES)~\cite{van2016x,bergmann2009x}, its inelastic scattering pathways give the resonant inelastic x-ray scattering spectroscopy (RIXS)~\cite{RevModPhys.73.203,RevModPhys.83.705,RevModPhys.93.035001}, and its Auger decay pathways give the x-ray photoemission spectroscopy (XPS)~\cite{RevModPhys.93.035001,lindau1974x,trinter2014resonant}, therefore the emerging of the cavity effect in x-ray regime with x-ray spectroscopy techniques might provide potential applications for x-ray core level spectroscopies.

\section{V. Conclusion}
In summary, we extended the semi-classical matrix and the quantum optical Green's function models into the platform of the x-ray thin-film planar cavity with the inner-shell electronic resonances. Firstly the classical Parratt's model is recalled, which is a standard method to calculate the reflectivity spectrum of the multilayer and used as benchmark for the matrix and Green's function models. Both of the semi-classical and quantum optical models can be successfully implemented to describe the recent experimental observations~\cite{Haber2019,PhysRevResearch.3.033063}, including the cavity enhanced decay rate, the cavity induced energy shift, the enhanced resonant elastic x-ray scattering and the multi-pathway Fano interference. Specifically, the semi-classical matrix model employs the ultrathin film expansion approximation, gives an analytical formula of the reflection coefficient, whose numerical spectra agree well with the ones of Parratt's calculation, and can give the interpretation of two-pathway interference. The CER and CIS are phenomenologically connected with the real and imaginary parts of the cavity field amplitude respectively. In the Green's function model, the classical field is firstly translated into the quantum field, then the system Hamiltonian is derived. The decay rate and spin-exchange of the atomic ensemble are obtained after considering the thin-film cavity as a 1D structural media. Based on the Hamiltonian and the Lindblad operator of the cavity system, it is successful to drive the effective energy-levels. The quantum Green's function model indicates that the cavity effect finally acts on the regulation of the intermediated core-hole state, so the cavity effect with electronic resonances would be very useful for the core level spectroscopy, and some possible applications are expected. Moreover, based on the quantum Green's function model the modular square of the dipole moment of the white line transition around L$_{\textrm{III}}$ edge of WSi$_2$ is obtained through the fitting process. The numerical spectra show very good agreement with the Parratt's results, which demonstrates the validity of the quantum Green's function model. On the other hand, because the slice approach for the atomic layer can be applied in the quantum Green's function model, the thin-film approximation can be always fulfilled. Therefore, the quantum Green's function model should be also suitable for the periodical multilayer structure which is normally used in soft and tender x-ray regimes. The two models of semi-classical matrix and quantum Green's function developed here would be helpful to predict the new cavity phenomena in the x-ray regime and promote the emerging between the quantum optical effects and the x-ray spectroscopy techniques.

%\end{spacing}

\section{Acknowledgements}
We gratefully acknowledge useful discussions with X. Kong, J. Liu, Z. Ma, W. Li, W. Xu, X. Liu, Y. Zhang and W. Xu. This work is supported by the National Natural Science Foundation of China (Grant No. U1932207), Strategic Priority Research Program of the Chinese Academy of Science (Grant No. XDB34000000), and the National Key Research and Development Program of China (Grant No. 2017YFA0402300). The financial support from the Heavy Ion Research Facility in Lanzhou (HIRFL) is also acknowledged.

%\bibliography{bib}
%===============================================================
%===============================================================

\end{document}